\documentclass[aps,pre,groupedaddress,twocolumn,10pt, reprint,superscriptaddress, longbibliography]{revtex4-2}

\usepackage[dvipsnames]{xcolor}
\usepackage{graphicx} 
\usepackage{mathtools}
\usepackage{amsfonts}
\usepackage{float}
\usepackage{svg}
\usepackage{hyperref}
\usepackage[normalem]{ulem}
\usepackage{url}
\usepackage{mathrsfs}
\usepackage{amsmath}
\usepackage{lipsum}
\usepackage{bm}

\allowdisplaybreaks[1]

\hypersetup{
    colorlinks=true,
    linkcolor={PineGreen},
    citecolor={PineGreen},
    urlcolor={PineGreen}
}

    \makeatletter
\def\@fnsymbol#1{\ensuremath{\ifcase#1\or \dagger\or \ddagger\or
   \mathsection\or \mathparagraph\or \|\or **\or \dagger\dagger
   \or \ddagger\ddagger \else\@ctrerr\fi}}
    \makeatother

\begin{document}

\title{Active Brownian particle under stochastic orientational resetting}

\author{Yanis Baouche}
\affiliation{Max Planck Institute for the Physics of Complex Systems, N\"othnitzer Stra{\ss}e 38,
01187 Dresden, Germany}
\author{Thomas Franosch}
\affiliation{Institut für Theoretische Physik, Universität Innsbruck, Technikerstra{\ss}e 21A, A-6020 Innsbruck, Austria}
\author{Matthias Meiners}
\affiliation{Mathematisches Institut, Justus-Liebig-Universität Gie{\ss}en,
Arndtstraße 2, 35392 Gießen, Germany}
\author{Christina Kurzthaler}
\email{ckurzthaler@pks.mpg.de}
\affiliation{Max Planck Institute for the Physics of Complex Systems, N\"othnitzer Stra{\ss}e 38,
01187 Dresden, Germany}
\affiliation{Center for Systems Biology Dresden, Pfotenhauerstra{\ss}e 108, 01307 Dresden, Germany}
\affiliation{Cluster of Excellence, Physics of Life, TU Dresden, 01307 Dresden, Germany}

\begin{abstract}
We employ renewal processes to characterize the spatiotemporal dynamics of an active Brownian particle under stochastic orientational resetting. By computing the experimentally accessible intermediate scattering function (ISF) and reconstructing the full time-dependent distribution of the displacements, we study the interplay of rotational diffusion and resetting. The resetting process introduces a new spatiotemporal regime reflecting the directed motion of agents along the resetting direction at large length scales, which becomes apparent in an imaginary part of the ISF. We further derive analytical expressions for the low-order moments of the displacements and find that the variance displays an effective diffusive regime at long times, which decreases for increasing resetting rates. At intermediate times the dynamics are characterized by a negative skewness as well as a non-zero non-Gaussian parameter. 
\end{abstract}

\maketitle

\section{Introduction}

Many microorganisms employ swimming mechanisms to optimize their survival strategies and explore their diverse habitats, where they are subject to strong confinements, hydrodynamic flows, and various external stimuli~\cite{Bell1990,Viswanathan2011, Goldstein2015, Gilpin2020,Thornton2020, Abe2020}. Environmental cues can steer the orientation of microswimmers and thus generate interesting dynamics emerging from the translational-orientational coupling. Chemotaxis, the movement along chemical gradients, has been the most extensively studied one so far~\cite{Wadhams2004}, but there is a plethora of directed movements along external stimuli, ranging from moisture cues (hydrotaxis)~\cite{Pringault2004} and sound cues (phonotaxis)~\cite{Nolen1986, Pollack2010} to magnetic fields (magnetotaxis) \cite{Lins2009,Painter2019}. Microswimmer properties, such as shape and density, can even lead to motion against gravity (gravitaxis)~\cite{Hader1999, Hagen2014, Hader2017, Chepizhko2022} and applied flows (rheotaxis)~\cite{Montgomery1997, Morales2015, Baker2019}. Besides external fields, bacteria can self-generate chemical and electrical signals, coordinating collective behaviors~\cite{Miller2001, Waters2005} and the formation of biofilms~\cite{Humphries2017}. 

In recent years, man-made agents have been engineered in the lab, which allows reproducing biological behaviors at the mesoscopic scale~\cite{Nelson2008, Solovev2012, Safdar2018}. Alongside the recent developments in micro-robots engineering~\cite{Nauber2023, Liu2023, Aziz2021, Aziz2022, Soto2020}, the question of characterizing the dynamics of chemically controlled swimmers in complex media is becoming more and more relevant. Increased control over chemical agents has been recently demonstrated in experiments in which tunable rotational diffusivity was prescribed~\cite{Rodriguez2020} or in which magnetic fields and acoustic waves were used to remotely control their speed and swimming orientation~\cite{Deng2023, Zhang2023}.

To describe the search strategies of those complex agents, the framework of stochastic resetting has been developed~\cite{Evans2020}. Applied in this context, resetting an agent to a certain prescribed position has proven to be an advantageous approach to reach a target, whether for a Brownian particle~\cite{Evans2011,Reuveni2016} or for an active particle \cite{Sar2023} in a free environment, and can introduce interesting non-equilibrium physics in the presence of external forces~\cite{rayDiffusionResettingLogarithmic2020, abdoliStochasticResettingActive2021}.
The dynamics of an active particle that stochastically resets its swimming direction have received less attention and so far provided analytical results in some parameter and temporal regimes using perturbative approaches~\cite{Kumar2020}, yet a characterization of the full spatiotemporal dynamics has not been elaborated. 

Here, we propose a minimal model to describe analytically such processes and to provide a framework for the characterization of its dynamics. We first formulate the orientational stochastic resetting as a renewal process, allowing us to obtain an equation for the spatiotemporal probability distribution. We then study the interplay of resetting and rotational diffusion, starting with analytical expressions of the low-order moments of the displacements. Subsequently, we calculate and discuss the characteristic function, also known as the intermediate scattering function (ISF), and finally the probability distribution is resolved for a wide range of length and time scales. 

\section{Theory}
In this section we introduce the model for the microscopic dynamics of an active Brownian particle (ABP) under orientational stochastic resetting and derive the corresponding renewal equations for the probability densities in Subsec.~\ref{sec:renewal}. We then present the computation of the intermediate scattering function and the low-order moments in Subsecs.~\ref{sec:ISF}-\ref{sec:moments}. 

\subsection{Model\label{sec:model}}

\begin{figure}
\centering
\includegraphics[width=\columnwidth]{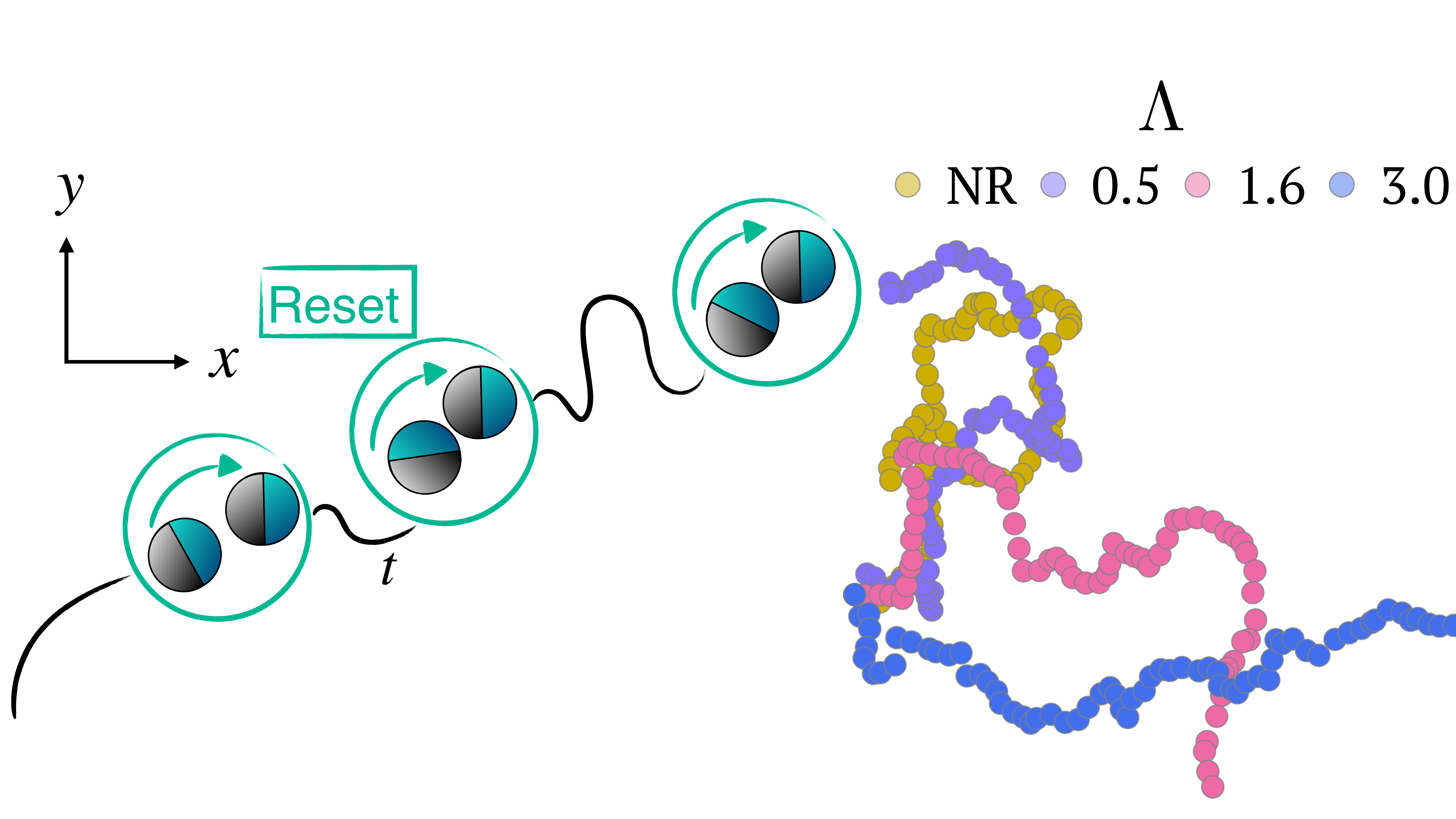}
\caption{({\it Left panel}) Schematic of the motion of an ABP under stochastic orientational resetting with distribution $\mathcal{P}(\vartheta)=\delta(\vartheta-\vartheta_r)$. ({\it Right panel}) Trajectories for different reduced resetting rates $\Lambda = \lambda/D_{\rm rot}$. Here, NR corresponds to the non-resetting case and the resetting events are distributed exponentially, $T(t) = \lambda e^{-\lambda t}$. The P\'eclet number is $\mathrm{Pe} = 10$.}
\label{schematic}
\end{figure} 


We consider an ABP in a two dimensional plane, moving at a constant speed~$v$ along its instantaneous orientation $\bm{u}(\vartheta) = (\cos(\vartheta), \sin(\vartheta))$~\cite{Hagen2011, Romanczuk2012, Kurzthaler2018}, see Fig.~\ref{schematic}~({\it left panel}). The particle displays translational and rotational diffusion with diffusion coefficients $D$ and $D_{\mathrm{rot}}$, respectively. Furthermore, the orientation $\vartheta(t)$ is reset to the resetting direction $\vartheta$  drawn from a distribution with probability density $\mathcal{P}(\vartheta;\vartheta_r)$ with mean $\vartheta_r$. In the following and without loss of generality, we set $\vartheta_r=0$ and hence $\mathcal{P}(\vartheta;\vartheta_r)\equiv\mathcal{P}(\vartheta)$. The resetting occurs at random times drawn from a distribution with probability density $T(t)$ of finite mean $\tau$. 
The particle's displacement $\bm{r}$ and change of orientation $\vartheta$ can be described by the following set of stochastic differential equations
\begin{subequations}
\begin{align}
    \dot{\bm{r}}(t) &= v \bm{u}(\vartheta) + \sqrt{2D} \boldsymbol{\eta}(t), \label{stoch_r}\\ 
    \dot{\vartheta}(t) &= \sqrt{2D_{\mathrm{rot}}} \xi (t), \label{stoch_theta}\\  
    \vartheta(t) & \to \hat{\vartheta} \text{ with } T(t) \text{ and } \hat{\vartheta}\sim\mathcal{P}(\hat\vartheta),
\end{align}
\end{subequations}

\noindent where $\to$ assigns an angle $\hat{\vartheta}$ to $\vartheta (t)$ and $\sim$ denotes that $\hat{\vartheta}$ is distributed according to $\mathcal{P}(\hat{\vartheta})$. Further, $\boldsymbol{\eta}(t)$ and $\xi(t)$ are independent Gaussian white noises of zero mean and delta correlated variance, $\langle \xi(t) \xi(t') \rangle = \delta(t-t')$ and $\langle \eta_i(t) \eta_j(t') \rangle = \delta_{ij} \delta(t-t')$ for $i,j = 1,2$. To describe the process, we introduce $\tau_{\mathrm{rot}} = D_{\mathrm{rot}}^{-1}$ as characteristic time scale and define the characteristic length scale by $a = \sqrt{3D/(4D_{\mathrm{rot}})}$, corresponding to the radius of a spherical particle under equilibrium diffusion coefficients. Two dimensionless parameters emerge: the P\'eclet number $\mathrm{Pe} = va/D$ characterizes the strength of active motion with respect to the diffusive processes. In addition, the reduced resetting rate $\Lambda = 1/ (\tau D_\mathrm{rot})$ measures the prevalence of resetting events compared to rotational diffusion. Finally, we note that the persistence length $L= v/ D_\mathrm{rot}$ of an ABP is the length travelled before it looses its orientation due to rotational diffusion.

Examples of different trajectories are displayed in Fig.~\ref{schematic}~({\it right panel}) and show that for increasing $\Lambda$, the agents travel a longer distance along the resetting direction, as the particle does not have time to reorient through rotational diffusion between two resetting events. We now derive renewal equations to describe the process and analytically compute statistical quantities.

\subsection{Renewal equations \label{sec:renewal}}

\begin{figure}[htp]
\centering
\includegraphics[width=\columnwidth]{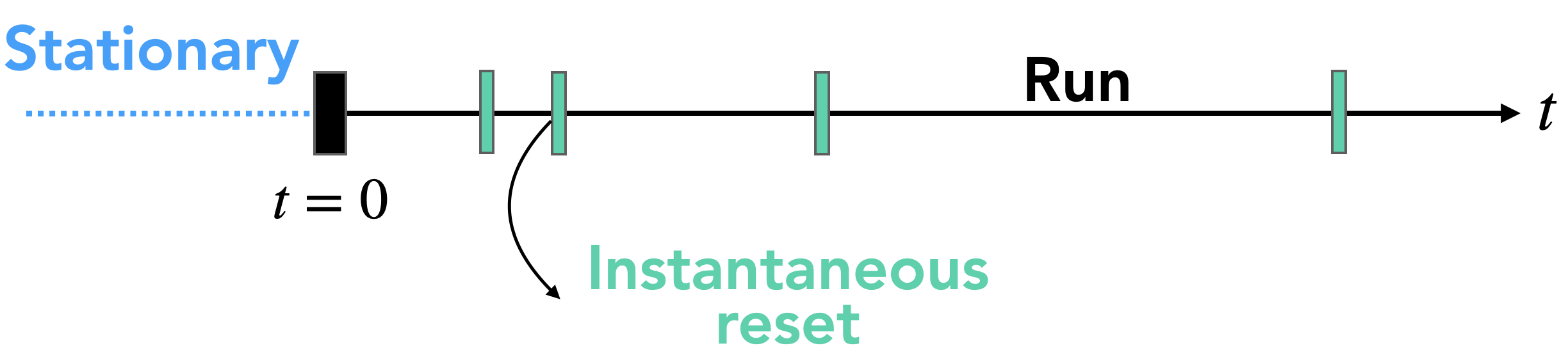}
  \caption{Schematic of the renewal process in time $t$, which denotes the lag time between the initial time of observation ($t=0$) and the current time. \label{fig:evolve}}
\end{figure} 

We describe the dynamics of the agent as a renewal process, which starts after the agent has already evolved a certain amount of time and is in a non-equilibrium stationary state for the displacement and the angle (see Fig.~\ref{fig:evolve}). Therefore, we introduce the lag time~$t$ between the initial time where we start to observe the particle and the current time. This aspect is particularly important for experiments where agents have already moved prior to the measurement. In practice, at $t=0$ the agent does not start with a fixed (resetting) orientation and the stationary distribution of the angle needs to be computed.

We denote by $P_n(\bm{r},t)$ the probability of the agent to have displaced a distance $\bm{r}$ during lag time $t$ and having undergone precisely $n\in\mathbb{N}_{0}$ resetting events. Similarly, $Q_n(\bm{r},t)$ is the probability density per unit time for the $n$-th resetting event ($n\in\mathbb{N}$) at time $t$ while having displaced $\bm{r}$, assuming that it does not depend on the agent's current (or future) orientation $\vartheta$.
Furthermore, we need to specify the dynamics between two resetting events and assume that the agent moves as an ABP with propagator $\mathbb{P}_{\mathrm{ABP}}(\bm{r}, \vartheta, t| \vartheta_0)$, that is, the conditional probability density for having displaced $\bm{r}$ in time $t$ and displaying an angle $\vartheta$ at lag time $t$ provided the initial orientation at time $0$ is $\vartheta_{0}$. Thus the product $\mathbb{P}_{\mathrm{ABP}}(\bm{r}, \vartheta, t| \vartheta_0) \mathcal{P}(\vartheta_0)$ represents the probability density to displace a distance $\bm{r}$ during lag time $t$ with final angle $\vartheta$, starting from an angle $\vartheta_0$ with a probability given by $\mathcal{P}(\vartheta_0)$. It is convenient to introduce the probability density for the agent to displace $\bm{r}$ during $t$ by
\begin{align}
    \mathbb{P}(\bm{r}, t) = \int_0^{2\pi} \! \mathrm{d}\vartheta \int_0^{2\pi} \! \mathrm{d}\vartheta_0 \, \mathbb{P}_{\mathrm{ABP}}(\bm{r}, \vartheta, t| \vartheta_0)\mathcal{P}(\vartheta_0), \label{eq:p}
\end{align}
where we have averaged over the initial orientation $\vartheta_0$ and integrated over $\vartheta$. In this work,  we choose the agent's orientation to be reset exactly to the $(Ox)$ axis $\mathcal{P}(\vartheta_0) = \delta(\vartheta_0)$, which allows simplifying Eq.~\eqref{eq:p} to $\mathbb{P}(\bm{r},t) = \mathbb{P}_{\mathrm{ABP}}(\bm{r}, t| \vartheta_0=0)$.

For $n\geq1$ the probability per unit time that the $(n+1)$-th resetting event takes place depends on the last resetting event $Q_n(\bm{r}-\bm{l},t-t')$ at a previous displacement $\bm{r} - \bm{l}$ and lag time $t-t'$. It is obtained via integration over all possibles displacements and lag times:
\begin{align}
    Q_{n+1}(\bm{r},t) = \int_{\mathbb{R}^2}\! \mathrm{d} \bm{l} \int_0^t \! \mathrm{d}t' Q_n(\bm{r}-\bm{l},t-t') T(t') \mathbb{P}(\bm{l}, t'). \label{eq:qn}
\end{align}
The integral equation for $P_n(\bm{r},t)$ can be derived from $Q_n(\bm{r},t)$:
\begin{equation}
    P_n(\bm{r},t) = \int_{\mathbb{R}^2} \!\mathrm{d} \bm{l} \!\int_0^t \! \mathrm{d}t' Q_n(\bm{r}-\bm{l},t-t') S(t') \mathbb{P}(\bm{l}, t'),\label{eq:pn}
\end{equation}
where $S(t)= \int_{t}^\infty  \mathrm{d}u T(u)$ is the survival probability (i.e. the probability that the resetting time exceeds $t'$). The interpretation of Eq. \eqref{eq:pn} is the following: the probability to displace a distance $\bm{r}$ at lag time $t$ is the sum over all possibilities to displace a distance $\bm{l}$ at lag time $t'$ given a previous reset at point $(\bm{r}-\bm{l}, t-t')$ in the absence of resetting during $[t-t',t]$. Then, the probability of the agent to displace a distance $\bm{r}$ at lag time~$t$ is obtained by summing over all resetting events: $P(\bm{r},t)=\sum_{n=0}^{\infty} P_n(\bm{r},t)$. Similarly, the rate of resetting events at displacement $\bm{r}$ and lag time $t$ is $Q(\bm{r},t) = \sum_{n=1}^{\infty} Q_n(\bm{r},t)$. 

To complete the previous equations, we need to determine $Q_{1}(\bm{r},t)$, the probability density per unit time for the first resetting event, as well as $P_{0}(\bm{r},t)$, the probability for the motion without resetting events. If we let $T^{1}(t)$ be the probability density for the first resetting event and $\mathbb{P}^0(\bm{r},t)$ the probability density to find the agent displace $\bm{r}$ at lag time $t$ in the non-equilibrium stationary state for the displacement and orientation, we find:

\begin{subequations}
\begin{align}
Q_1(\bm{r}, t) &= \mathbb{P}^0(\bm{r},t)T^1(t), \\ 
P_0(\bm{r}, t) &=\mathbb{P}^0(\bm{r},t)\int_t^{\infty}\mathrm{d}t' T^1(t'). \label{eq:p0}
\end{align}
    
\end{subequations}

Thus, after summation, Eqs.~\eqref{eq:qn}-\eqref{eq:pn} become~\cite{Cox1962, Feller1991, Angelani2013, Zaburdaev2015, Kurzthaler2024, Zhao2024} 
\begin{subequations}
\begin{align}
\begin{split}
Q(\bm{r}, t) &= Q_1(\bm{r}, t) +\\ 
& \int_{\mathbb{R}^2} \mathrm{d} \bm{l} \int_0^t \mathrm{d}t' Q(\bm{r}-\bm{l},t-t') T(t')\mathbb{P}(\bm{l},  t'),   
 \label{renew_q}
 \end{split}\\
 \begin{split}
P(\bm{r}, t) &= P_0(\bm{r}, t) +\\ 
&\int_{\mathbb{R}^2} \mathrm{d} \bm{l} \int_0^t \mathrm{d}t'  Q(\bm{r}-\bm{l},t-t') S(t') \mathbb{P}(\bm{l},t'). \label{renew_p}
 \end{split}
\end{align}
\end{subequations}

It can be shown that $T^1(t)=S(t)/\tau$, which assures that in the stationary state the probability for a renewal event is independent of time $t$ and amounts to $1/\tau$ (see Appendix \ref{evolving_phase_deriv}).
Furthermore, we note that $\mathbb{P}^0(\bm{r},t)$  is given by 
\begin{equation}
	\mathbb{P}^0(\bm{r},t) = \int_0^{2\pi} \mathrm{d}\vartheta\int_0^{2\pi} \mathrm{d}\vartheta_0 \ \mathbb{P}_{\mathrm{ABP}}(\bm{r},\vartheta,t | \vartheta_0)p^0(\vartheta_0),
\end{equation}
where $p^0(\vartheta_0)$ is the stationary distribution of the orientation before a resetting event. The latter depends on the distribution of resetting events via (see Appendix~\ref{evolving_phase_deriv})
\begin{align}
p^0(\vartheta) &= \frac{1}{\tau} \int_0^\infty \mathrm{ d}t \ S(t)\mathbb{P}(\vartheta,t),
\end{align}
where $\mathbb{P}(\vartheta,t)$ denotes the probability distribution of the orientation. 

We note that $p^0$ has previously been computed for the case of a Brownian particle resetting its {\it position} to  $0$ at exponentially distributed times with rate~$\lambda$~\cite{Evans2011}: 
\begin{align}
p^0_{\mathrm{pos}}(\vartheta) = \frac{\sqrt{\Lambda}}{2}  e^{-\sqrt{\Lambda}|\vartheta|},\label{eq:stationary_BP}
\end{align}
where $\vartheta$ corresponds to the position $x \in \mathbb{R}$ and $\Lambda$ to $\lambda/D$ with translational diffusion coefficient $D$. 

In what follows, we choose exponentially distributed resets, $T(t)=\lambda e^{-\lambda t}$, for our process 
Since in our case the orientation is reset instead of the particle's position, we need to wrap Eq.~\eqref{eq:stationary_BP} around $-\pi$ to $\pi$, resulting in a closed expression 
\begin{align}
\label{closed_p0}
p^0(\vartheta) = \frac{\sqrt{\Lambda}}{2} \frac{\cosh(\sqrt{\Lambda}(\pi-|\vartheta|))}{\sinh(\sqrt{\Lambda}\pi)},
\end{align}
where $D$ in Eq.~\eqref{eq:stationary_BP} is replaced by the rotational diffusion coefficient $D_{\rm rot}$. See Appendix~\ref{evolving_phase_deriv} for a derivation. 

\subsection{Intermediate scattering function \label{sec:ISF}}

The renewal equations (Eqs.~\eqref{renew_q}-\eqref{renew_p}) are coupled and involve two convolution products. Performing a spatial Fourier transform

\begin{equation}
    \widetilde{P}(\bm{k},t) = \int_{\mathbb{R}^{2}} \mathrm{d} \bm{r}  P(\bm{r},t) e^{- i \bm{k} \cdot \bm{r}},
\end{equation}
we arrive at
\begin{subequations}
\begin{align}
	\widetilde{Q}(\bm{k}, t) &= \widetilde{Q}_1(\bm{k}, t) + \int_0^t \mathrm{d}t' \widetilde{Q}(\bm{k},t-t') T(t')\widetilde{\mathbb{P}}(\bm{k}, t'), \label{re_fe_q} \\
	\widetilde{P}(\bm{k}, t) &= \widetilde{P}_0(\bm{k}, t) + \int_0^t  \mathrm{d}t' \widetilde{Q}(\bm{k},t-t') S(t') \widetilde{\mathbb{P}}(\bm{k}, t'), \label{re_fe_p}
\end{align}
\end{subequations}
and note that $\widetilde{\mathbb{P}}(\bm{k}, t')$ can be expressed analytically in terms of Mathieu functions~\cite{Kurzthaler2018}, see Eq.~\eqref{ISF_free} in Appendix~\ref{ISF_ABP_NU}. These coupled equations can thus be solved numerically for each $\bm{k} = (k_x,k_y)$ separately. 

We note that $\widetilde{P}(\bm{k},t)$ represents the characteristic function of the stochastic process, also known as the intermediate scattering function (ISF)
which measures the dynamics at length scales $2\pi/ |\bm{k}|$ along directions $\bm{k}/|\bm{k}|$ at lag time $t$ ($\Delta \bm{r}(t) = \bm{r}(t) - \bm{r}_{0}$). It encodes the moments of the displacements via a small-wavenumber expansion: 
\begin{equation}\label{expansion_isf_time}
\begin{split}
\widetilde{P}(\bm{k}, t) &=  \biggl \langle e^{-i \bm{k} \cdot \Delta \bm{r}(t) } \biggl \rangle \\ 
&= 1 - ik_x \langle \Delta x(t) \rangle - ik_y \langle \Delta y(t) \rangle \\ 
&-  \frac{k_x^2}{2} \langle \Delta x^2(t) \rangle -  \frac{k_y^2}{2}\langle \Delta y^2(t) \rangle \\ 
&- k_x k_y \langle \Delta x(t) \Delta y(t)\rangle  + \mathcal{O}(\lvert \bm{k} \rvert^{3}).
\end{split}
\end{equation}

\subsection{Derivation of the moments\label{moment_deriv} \label{sec:moments}}
To obtain analytical expressions for the low-order moments, we take the Laplace transform 

\begin{equation}
    \widehat{P}(\bm{k},s) = \int_{0}^{\infty} \mathrm{d} t  P(\bm{k},t) e^{-st}
\end{equation}
of Eqs.~\eqref{re_fe_q}-\eqref{re_fe_p}:
\begin{subequations}
\begin{align}
	\widehat{Q}(\bm{k}, s) &=  \frac{\widehat{Q}_1(\bm{k},s)}{1 - \mathcal{L}[T(t)\widetilde{\mathbb{P}}(\bm{k}, t)](s)}, \label{re_la_q} \\
	\widehat{P}(\bm{k}, s) &=  \widehat{P}_0(\bm{k},s) + \widehat{Q}(\bm{k},s) \mathcal{L}[S(t) \widetilde{\mathbb{P}}(\bm{k}, t)](s) \label{re_la_p}
\end{align}
\end{subequations}
with $\mathcal{L}[f(t)](s) = \int_0^{\infty} f(t) e^{-st} \mathrm{d}t$. Since the moments of the displacements in the run phase are known (see Appendix~\ref{appendix_abp_moments}), we can expand $\widetilde{\mathbb{P}}(\bm{k}, t)$ and compute the expansions of $\widetilde{Q}_1(\bm{k},t)$ and $\widetilde{P}_0(\bm{k},t)$. Inserting those expansions into Eqs.~\eqref{re_la_q}-\eqref{re_la_p} (in Laplace space) and expanding both sides in terms of $k_x$ and $k_y$ gives:
\begin{align}  \label{expansion_isf}
 \widehat{P}(\bm{k}, s) &= 1 - ik_x \mathcal{L}\Big[\langle \Delta x(t) \rangle\Big] (s) - ik_y \mathcal{L}\Big[\langle \Delta y(s) \rangle\Big] (s) \nonumber \\
&-  \frac{k_x^2}{2} \mathcal{L} \Big[\langle \Delta x^2(t) \rangle \Big](s) -  \frac{k_y^2}{2}  \mathcal{L} \Big[\langle \Delta y^2(t) \rangle \Big](s)  \\ 
&- k_x k_y   \mathcal{L} \Big[\langle \Delta x(t) \Delta y(t) \rangle \Big](s) + \mathcal{O}(\lvert  \bm{k} \rvert^{3}),  \nonumber
\end{align}
which allows us to identify the moments of the full process.  An inverse Laplace transform yields the moments in the time variable.

\begin{figure*}[tp]
\centering
  \includegraphics[width=\textwidth]{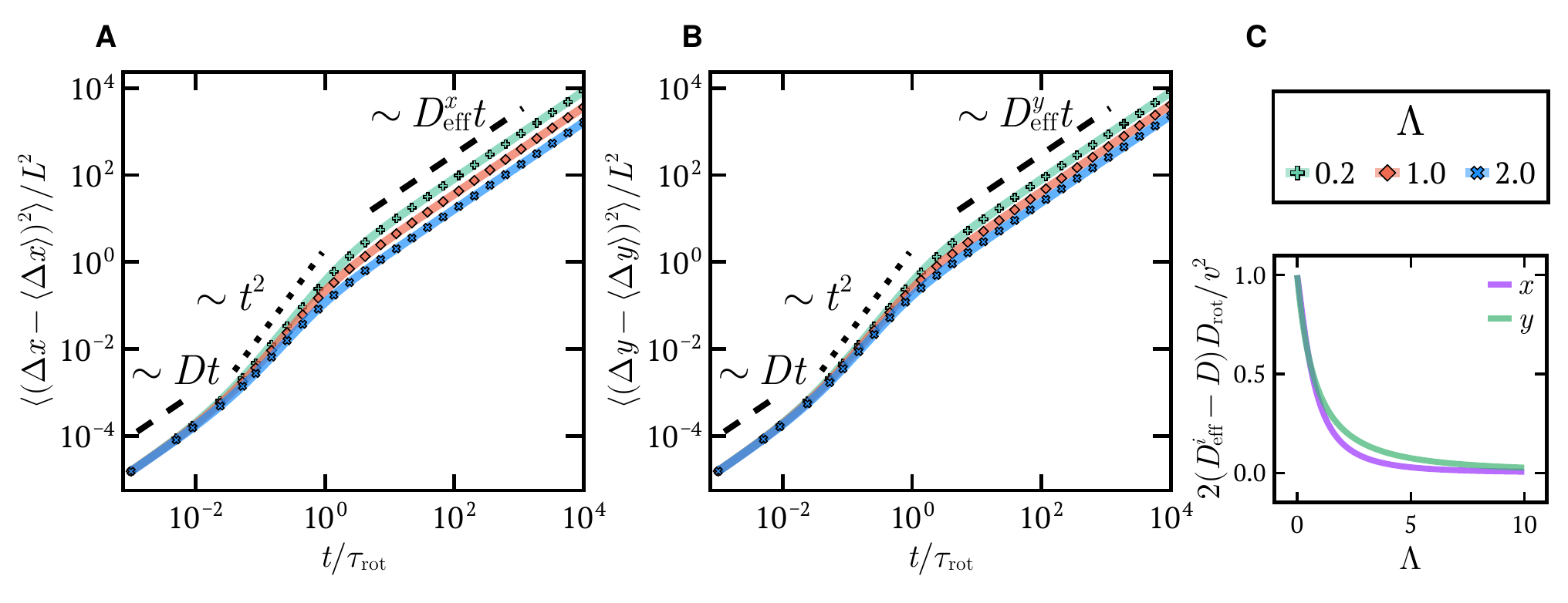}
     \caption{(\textbf{A-B}) Variance in the $x$ and $y$ directions as a function of time for different reduced resetting rates $\Lambda = \lambda/D_{\mathrm{rot}}$.  (\textbf{C}) Effective diffusivities as a function of $\Lambda$. The P\'eclet number is $\mathrm{Pe}=10$ and $\tau_{\rm rot}$ denotes the rotational relaxation time. Theory and simulation are shown with lines and symbols, respectively.}
    \label{variances}
\end{figure*}

\section{Results}
Here, we discuss the moments of the displacements, including the mean motion, variances, skewness, and non-Gaussian parameter (Subsec.~\ref{results:moments}). We then analyze the different spatiotemporal regimes of the ISF in Subsec.~\ref{results:ISF} and subsequently resolve the full probability distribution of the displacements in Subsec.~\ref{results:dist}. We show results for exponentially distributed resets, i.e. $T(t) = \lambda e^{-\lambda t}$, a resetting distribution $\mathcal{P}(\vartheta)=\delta(\vartheta)$, and a fixed P{\'e}clet number, ${\rm Pe}=10$, unless otherwise stated.

\subsection{Low-order moments \label{results:moments}}
The resetting mechanism introduces an overall drift of the active agents along the resetting direction
\begin{equation}
    \langle \Delta \bm{r}(t) \rangle = v t  \frac{\Lambda}{1+\Lambda} \bm{e}_x.
    \label{eq_mean_disp}
\end{equation}
For $\Lambda\ll1$, the drift grows linearly in the reduced resetting rate $\langle \Delta \bm{r}(t) \rangle \simeq v t  \Lambda\bm{e}_x$, while for $\Lambda\gg1$ we find directed motion along the resetting direction $\bm{e}_x$, as frequent resetting events prevent the particle's orientation to diffuse. We also notice that the mean displacement depends linearly on time, which is consistent with the fact that the velocity has to be independent of time since we start in a stationary state for the displacement.

We next quantify the fluctuations around the mean drift in terms of the variance along and perpendicular to the resetting direction:

\begin{widetext}
\begin{subequations}
\begin{align}
    \Big\langle \big( \Delta x(t) - \langle \Delta x(t) \rangle \big)^2 \Big\rangle = 2Dt + \frac{2v^2}{D_{\mathrm{rot}}^2} \frac{2+5\Lambda}{(1+\Lambda)^4 (4+\Lambda)} \left[ e^{-tD_{\mathrm{rot}}(1+\Lambda)} -1 + tD_{\mathrm{rot}} (1+\Lambda) \right],
 \label{vx} \\ 
    \Big\langle \big( \Delta y(t) - \langle \Delta y(t) \rangle \big)^2 \Big\rangle = 2Dt + \frac{2v^2}{D_{\mathrm{rot}}^2} \frac{2}{(1+\Lambda)^2 (4+\Lambda)} \left[ e^{-tD_{\mathrm{rot}}(1+\Lambda)} -1 + tD_{\mathrm{rot}} (1+\Lambda) \right]. \label{vy}   
\end{align}
\end{subequations}
\end{widetext}
We observe that at short times, $t\lesssim D/v^{2}\equiv\tau_{\mathrm{ diff}}$ (transition from diffusive regime timescale), both variances increase linearly in  time $\propto Dt$, reflecting the translational diffusion, see Figs.~\ref{variances}{\bf A}-{\bf B}. At intermediate times $\tau_{\mathrm{ diff}} \lesssim t\lesssim \tau_{\rm rot}/(1+\Lambda)$, a persistent behavior $\propto t^2$ appears due to the active motion and resetting mechanism. At long times, $t\gtrsim \tau_{\rm rot}/(1+\Lambda)$ we observe a transition to an effective diffusive regime, where the variance grows linearly in time, with an effective diffusivity determined by the reduced resetting rate $\Lambda$ (see Fig.~\ref{variances}{\bf C}):
\begin{align}
    \frac{2D_{\mathrm{rot}}}{v^2} (D_{\rm eff}^{x}-D) &= \frac{4+10 \Lambda}{(1+\Lambda)^3 (4 +\Lambda)}, \label{deffx} \\ 
    \frac{2D_{\mathrm{rot}}}{v^2} (D_{\rm eff}^{y}-D) &= \frac{4}{(1+\Lambda)(4 +\Lambda)}.  \label{deffy}
\end{align}
In particular, the effective diffusivity along  the resetting direction ($x$) decreases faster than that of the perpendicular ($y$) direction. If the agent's orientation is more frequently reset to a certain direction, it is more likely to move in a straight line where fluctuations become less. At $\Lambda = 0$, both diffusivities approach that of a simple ABP, $D_{\mathrm{eff}}^{\rm ABP} = D + v^2/(2D_{\mathrm{rot}})$ \cite{Cates2013}, which is in agreement with Ref.~\cite{Kumar2020} for $D=0$. In addition, the original mean-square displacement is recovered from Eqs. \eqref{vx}-\eqref{vy}. At large~$\Lambda$, rotational diffusion is nullified and the effective coefficients reduce to that of a translationally diffusive particle, $D_{\rm eff}^{x,y} = D$.

\begin{figure*}[tp]
  \centering
    \includegraphics[width=\textwidth]
  {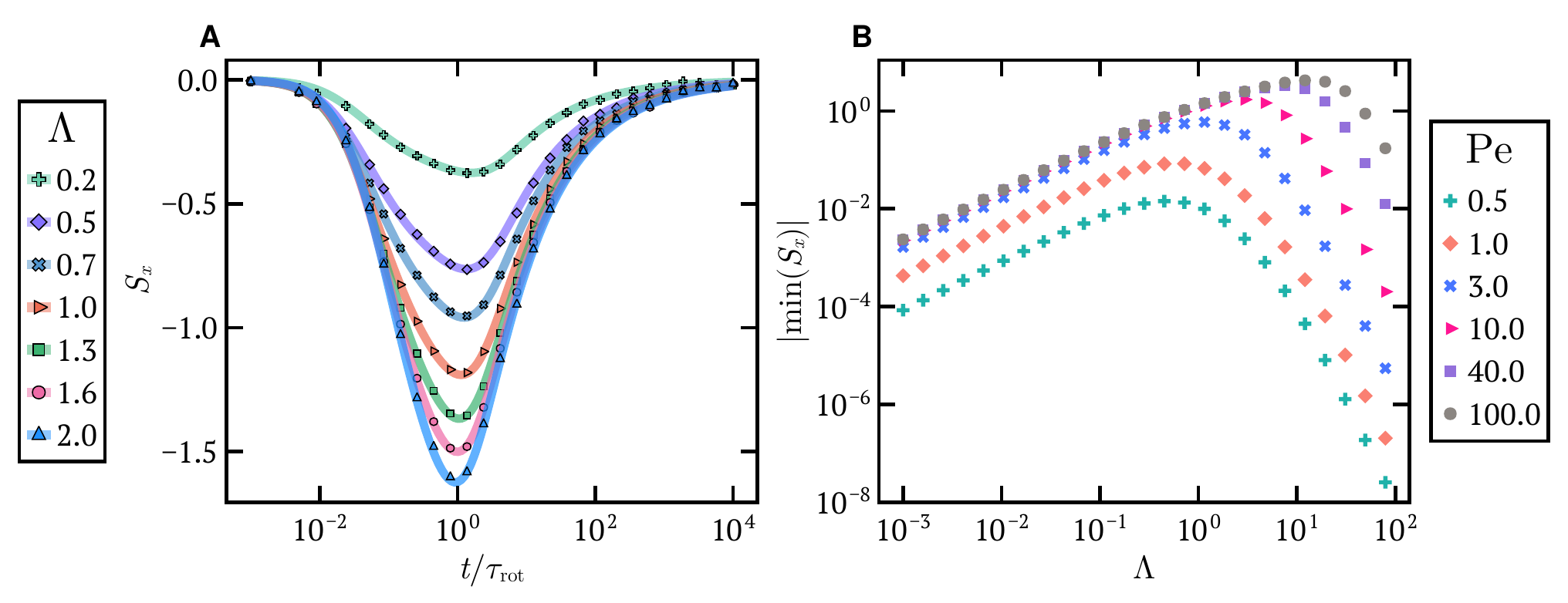} 
     \caption{(\textbf{A}) Skewness $S_{x}$ in the $x$-direction as a function of time for different reduced resetting rates $\Lambda = \lambda/D_{\mathrm{rot}}$. The P\'eclet number is $\mathrm{Pe}=10$ and $\tau_{\rm rot}$ denotes the rotational relaxation time. Theory and simulation are shown with lines and symbols, respectively. (\textbf{B})~Absolute value of the minimum of the skewness $S_{x}$ as a function of the reduced resetting rate $\Lambda$ for different P\'eclet numbers~$\mathrm{Pe}$.}
       \label{fig_skewness}
\end{figure*}

\begin{figure*}[tp]
\centering
    \includegraphics[width=\textwidth]{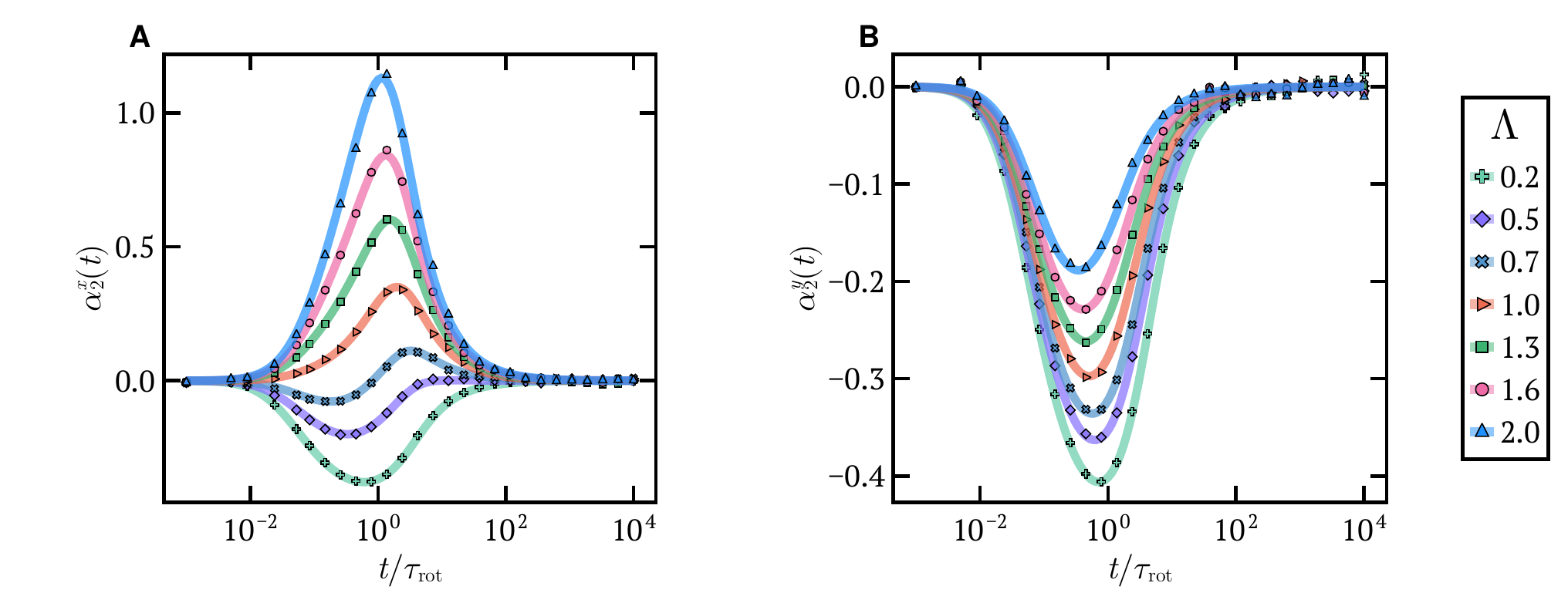}
     \caption{Non-Gaussian parameters, $\alpha_2^x(t)$ (\textbf{A}) and $\alpha_2^y(t)$ (\textbf{B}), in the $x-$ and $y-$directions for different reduced resetting rates $\Lambda = \lambda/D_{\mathrm{rot}}$. The P\'eclet number is $\mathrm{Pe}=10$ and $\tau_{\rm rot}$ denotes the rotational relaxation time. Theory and simulation are shown with lines and symbols, respectively.}
       \label{fig_ngp}
\end{figure*}

The contribution of the resetting mechanism also manifests itself in the velocity auto-covariance function \cite{HansenMcDonald2013}: 
\begin{equation}
    \begin{split}
     K(t) &= \left \langle \dot{\bm{r}}(t) \cdot \dot{\bm{r}}(0)  \right \rangle - \left \langle \dot{\bm{r}}(t) \right \rangle  \left \langle \dot{\bm{r}}(0) \right \rangle,\\ 
    &=  \frac{1}{2}\frac{\mathrm{d}^2}{\mathrm{d}t^2} \langle (\Delta \bm{r} (t) - \langle \Delta \bm{r} (t) )^2 \rangle, \\
    &= v^2 \frac{1+2\Lambda}{(1+\Lambda)^2} \exp \left( -\frac{1+\Lambda}{\tau_{\mathrm{rot}}}t \right).
    \end{split}
    \label{eq_autocov}
\end{equation}
We consider an auto-covariance function instead of an auto-correlation function because of the drift. Equation \eqref{eq_autocov} indicates a decay on the timescale $\tau_{\mathrm{rot}}/(1+\Lambda)$, which is faster than $\tau_{\mathrm{rot}}$ for a simple ABP, whose auto-correlation function reappears for $\Lambda=0$.

As a measure of the anisotropy of the resetting process, we consider the skewness along the resetting direction:
\begin{equation}
    S_x(t) = \frac{\langle (\Delta x (t) - \langle \Delta x (t) 
    \rangle)^3 \rangle}{\langle (\Delta x (t) - \langle \Delta x (t) \rangle )^2 \rangle^{3/2} }.
\end{equation}
It measures the asymmetry of the displacement distribution; in particular, a negative skewness indicates a distribution that is tilted to the right, or whose tail is on the left side. Given that we chose the probability density distribution $\mathcal{P}(\vartheta)$ to be symmetric,  the skewness vanishes along the $y-$direction.

Figure~\ref{fig_skewness}\textbf{A} shows the skewness for various $\Lambda>0$. It is negative at intermediate times and assumes a minimum, which decreases for increasing $\Lambda$. Furthermore, as $\Lambda$ is increased the minimum shifts to the left, reflecting that the impact of resetting occurs earlier. The negative skewness indicates that the distribution looses symmetry and is tilted towards the resetting direction. For small, $t\lesssim \tau_{\mathrm{diff}}$, and long times, $t\gtrsim \tau_{\rm rot}/(1+\Lambda)$, the skewness vanishes as the distribution distributions becomes Gaussian at those times.

Finally, we compute the non-Gaussian parameter (NGP)~\cite{Hofling2013}, defined as
\begin{subequations}
\begin{align}
    \alpha_2^{x}(t) &= \frac{1}{3}\frac{\langle (\Delta x - \langle \Delta x 
    \rangle)^4 \rangle}{\langle (\Delta x - \langle \Delta x \rangle )^2 \rangle^{2} } -1, \\
    \alpha_2^{y}(t) &= \frac{1}{3}\frac{\langle \Delta y ^4 \rangle}{\langle\Delta y^2 \rangle^{2} } -1,
\end{align} 
\end{subequations}
in the $x-$ and $y-$directions, respectively. It measures the existence of outliers/fat-tailed character of a distribution with respect to a normal law. 

Starting the analysis with the resetting ($x$) direction, Fig.~\ref{fig_ngp}\textbf{A} indicates that $\alpha_2^{x}(t)$ vanishes at short-times $t \lesssim \tau_{\mathrm{diff}}$ due to the diffusive motion. Two different behaviors are then to be observed, depending on the value of $\Lambda$. For $\Lambda \lesssim 1/2$, the NGP becomes negative because of the active motion, as the particle is mostly moving at speed~$v$. Since the resetting timescale $\tau = 1/\lambda$ is larger than the rotational diffusive timescale $\tau_{\mathrm{rot}}$, the resetting does not have the occasion to manifest itself and $\alpha_2^{x}$ eventually vanishes due to rotational diffusion. For large $\Lambda$, the agents have a preferred displacement direction and those reorienting due to rotational diffusion count as outliers and contribute to a positive NGP. The case $\Lambda = 0.7$ displays both a minimum and a maximum, demonstrating the intricate interplay of the two processes.

In the perpendicular ($y$) direction (see Fig.~\ref{fig_ngp}\textbf{B}), the NGP starts at $0$ because of translational diffusion and becomes negative due to the active motion at intermediate times. In particular, we observe a minimum at $t \simeq \tau_{\mathrm{rot}}/(1+\Lambda)$ that decreases as $\Lambda$ is decreased. This is due to the fact that less movement takes place in the perpendicular direction when the agent's orientation is increasingly reset. Once again, the position of the minimum shifts to shorter times as the resetting rate is increased. 

So far we have discussed the case of intermediate P\'eclet numbers ($\mathrm{Pe}=10$) and naturally the question arises as to how the results change with this number. While for the mean drift and the variance qualitative features do not change by varying the P\'eclet number, the skewness and non-Gaussian parameter are sensitive to it. We shed light on this by looking at the minimum of the skewness $S_{x}$ as a function of $\Lambda$ for different $\mathrm{Pe}$ (see Fig. \ref{fig_skewness} \textbf{B}). There is an optimal resetting rate $\Lambda_{0}$, such that (at intermediate times) the distribution is maximally skewed towards the resetting direction. Importantly, $\Lambda_{0}$ becomes greater as the P{\'e}clet number increases, which can be interpreted in the following way: if $\Lambda$ is very small, the agent can easily reorient and the distribution tends to be symmetric, if $\Lambda$ is excessively large, the agent simply displaces along the resetting direction and the skewness also vanishes. There is therefore a characteristic $\Lambda_0$ at which the agent can still reorient while mostly travelling towards the resetting direction. Similarly, one can rationalize the behavior of optima of the NGPs for varying P{\'e}clet numbers, which occur due to an optimal number of agents counting as outliers.




\begin{figure*}[tp]
  \centering
    \includegraphics[width=\textwidth]{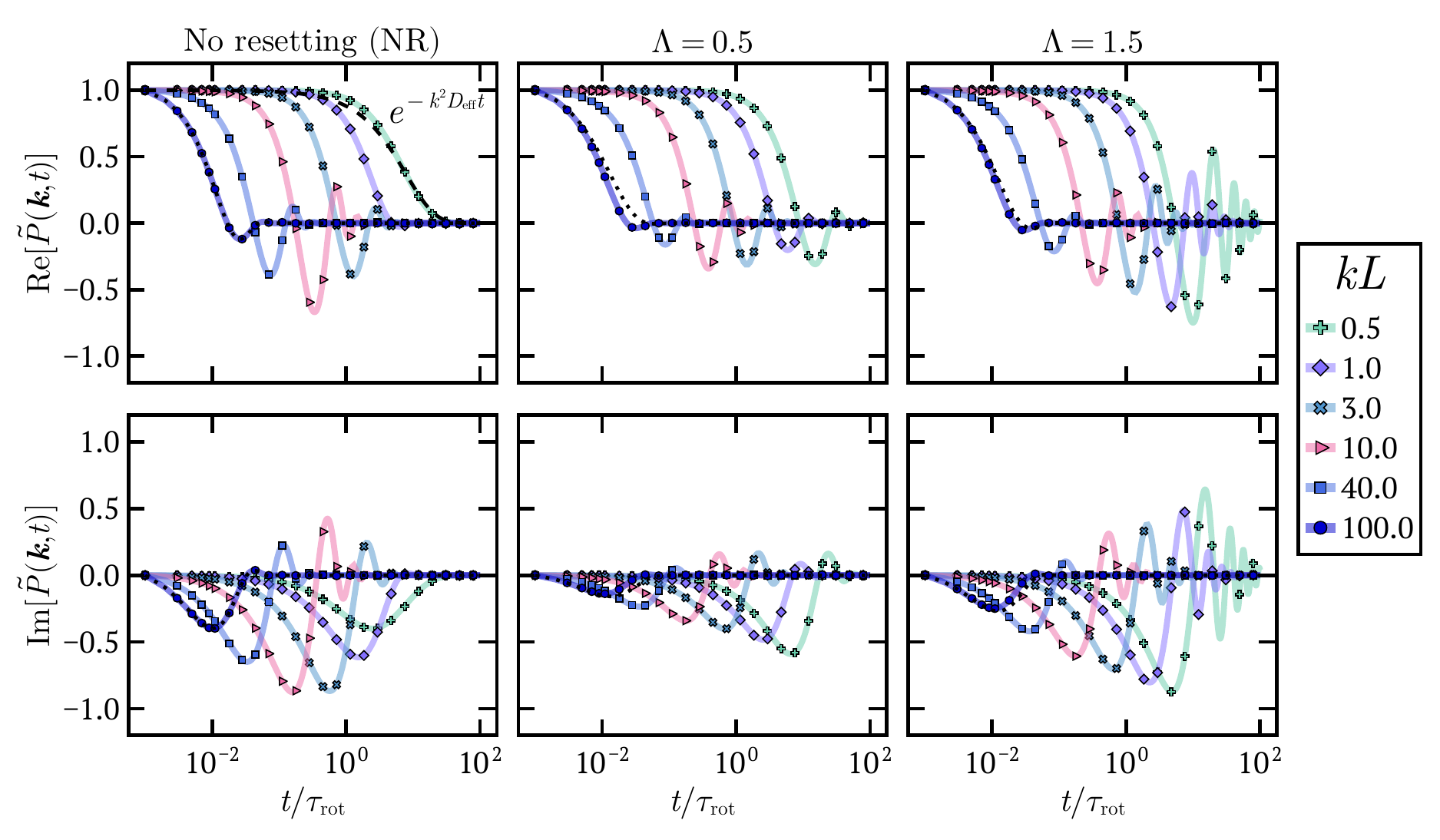}
      \caption{Intermediate scattering function $\widetilde{P}(\bm{k},t)$ on logarithmic time axis for several length scales $kL$ and reduced resetting rates $\Lambda = \lambda/D_{\mathrm{rot}}$. The P\'eclet number is $\mathrm{Pe}=10$ and $\tau_{\rm rot}$ denotes the rotational relaxation time. Theory and simulation are shown with lines and symbols, respectively. Black dotted lines correspond to the approximation $e^{-Dt k^2} e^{-i \bm{k} \cdot \langle \Delta \bm{r}(t)\rangle}$. The black dashed line corresponds to $e^{-k^{2}D_{\mathrm{eff}}t}$.}
  \label{isf_kl}
\end{figure*}

\subsection{Intermediate scattering function \label{results:ISF}}
We now resolve the ISF along the resetting direction, $\bm{k}/|\bm{k}|=\bm{e}_x$, for a large range of wavenumbers, $kL\in [0.5,100]$, and reduced resetting rates, $\Lambda$, and keep the P{\'e}clet number fixed, Pe=$10$. Therefore, we evaluate numerically Eqs.~\eqref{re_fe_q}-\eqref{re_fe_p} for each wave vector separately and compare our results with stochastic simulations (see Appendix~\ref{stochastic_simulations}).  

We first observe a non-vanishing imaginary part in all cases, showing the anisotropic character of the process due to the presence of a drift, see Fig.~\ref{isf_kl}~(lower row). In Eq.~\eqref{expansion_isf}, the odd moments are non-zero, giving rise to the imaginary part of the ISF. 

First, we discuss the ISF for an active agent that does not undergo any orientational resetting but starts to swim at a fixed initial orientation $\bm{u}(\vartheta_0)=\bm{e}_x$, see first column of Fig.~\ref{isf_kl}. The real and imaginary parts of the ISFs display strong oscillations due to the active movement at intermediate time, $\tau_{\rm diff}\lesssim t\lesssim \tau_{\rm rot}$, and length scales, $kL \simeq 2\pi$.  The oscillations smear out at short times, $t\lesssim \tau_{\rm diff}$, due to short-time translational diffusion. At large $kL$, the displacement of the agent is directed along its initial orientation and we can split the displacement into a deterministic part $\Delta \bm{r}(t)\simeq vt\bm{e}_x$ and fluctuations given by a relaxing exponential. As both processes are decoupled, the ISF can be approximated by $\widetilde{P}(\bm{k},t) = e^{-k^2Dt} \Big(\cos(\bm{k}\cdot \langle \Delta\bm{r}(t) \rangle)-i \sin(\bm{k}\cdot \langle \Delta\bm{r}(t) \rangle ) \Big)$ (see black dotted curves in Fig.~\ref{isf_kl}). This approximation is not appropriate at larger length scales, as it neglects orientational diffusion. For decreasing $kL$ the oscillations become smaller and vanish due to rotational diffusion at $t\gtrsim \tau_{\mathrm{rot}}$. At length scales larger than the persistence length, corresponding to $kL\lesssim 2\pi$, we observe a typical exponential decay $P(\bm{k},t) \propto e^{-k^2 D_{\mathrm{eff}}^{\rm ABP}t}$ accounting for an effective diffusive behavior. At these length scales the initial orientation of the agent does not affect the ISF and we recover the behavior of ABPs with a random initial orientation~\cite{Kurzthaler2018}. 

We observe two main differences in the presence of stochastic resetting compared to the non-resetting case. First, at small length scales $kL \gtrsim 2\pi$ the ISFs also display oscillations but their amplitude is lower than in the non-resetting (NR) case. This is due to the fact that the measurement starts when the agent's orientation is in a stationary state, in contrast to when the initial orientation is fixed. The second, more prominent difference lies in the behavior for large length scales $kL \lesssim 2\pi$ (see $kL=0.5$), where oscillations persist due to the resetting mechanism. 

The transition to large length-scale oscillations depends on the resetting rate $\Lambda$. For $\Lambda=0.5$, corresponding to $\lambda= 0.5D_{\mathrm{ rot}}$ (or $\tau=2.0\tau_{\rm rot}$), the swimming direction diffuses faster than the time it takes to be reset. Therefore, we observe oscillations at times $t\lesssim\tau_{\rm rot}$ and small length scales, which fade out at intermediate times $\tau_{\rm rot}\lesssim t\lesssim 1/\lambda$. At times $t\gtrsim 1/\lambda$ oscillations become larger, reflective of the directed motion due to the resetting mechanism. As the imaginary part is a fingerprint of the overall drift only, the oscillations grow as $kL$ increases. 

For $\Lambda=1.5$, corresponding to $\lambda=1.5D_{\mathrm{ rot}}$, the orientation of the swimmer is reset before it decorrelates due to rotational diffusion. Thus, we do not observe a damping of oscillations at intermediate times and length scales. Overall, the oscillations are larger than for $\Lambda=0.5$ due to the directed motion along the resetting direction. 

\begin{figure}[tp]
\centering
    \includegraphics[width=\columnwidth]{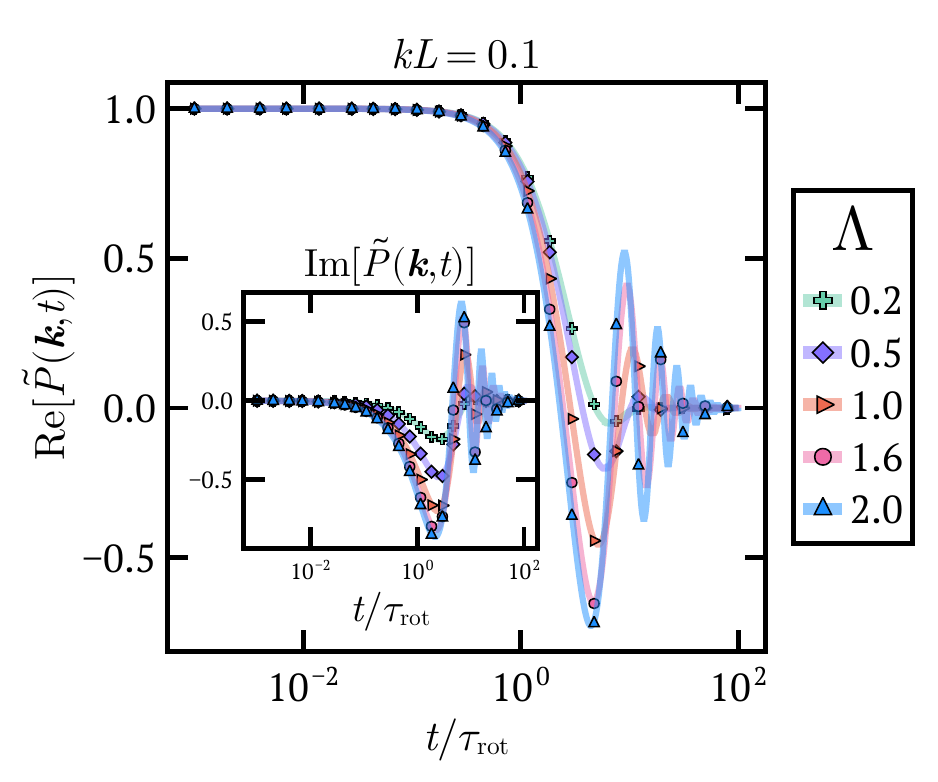}
    \caption{Intermediate scattering function $\widetilde{P}(\bm{k},t)$ for a fixed length scale $kL$ and several reduced resetting rates $\Lambda = \lambda/D_{\mathrm{rot}}$. The P\'eclet number is $\mathrm{Pe}=10$ and $\tau_{\rm rot}$ denotes the rotational relaxation time. Theory and simulation are shown with lines and symbols, respectively.}
      \label{ISF_lambda}
\end{figure}

One crucial difference of the ISFs for different $\Lambda$ lies in the large length scale $kL \lesssim 2\pi$ regime, see Fig. \ref{ISF_lambda} for $kL=0.1$. We observe that the reduced resetting rate influences the amplitude of the oscillations and characterizes the strength of this resetting-induced directed motion.

\subsection{Probability density distribution \label{results:dist}}
\begin{figure*}[tp]
  \centering
  \includegraphics[width=\textwidth]{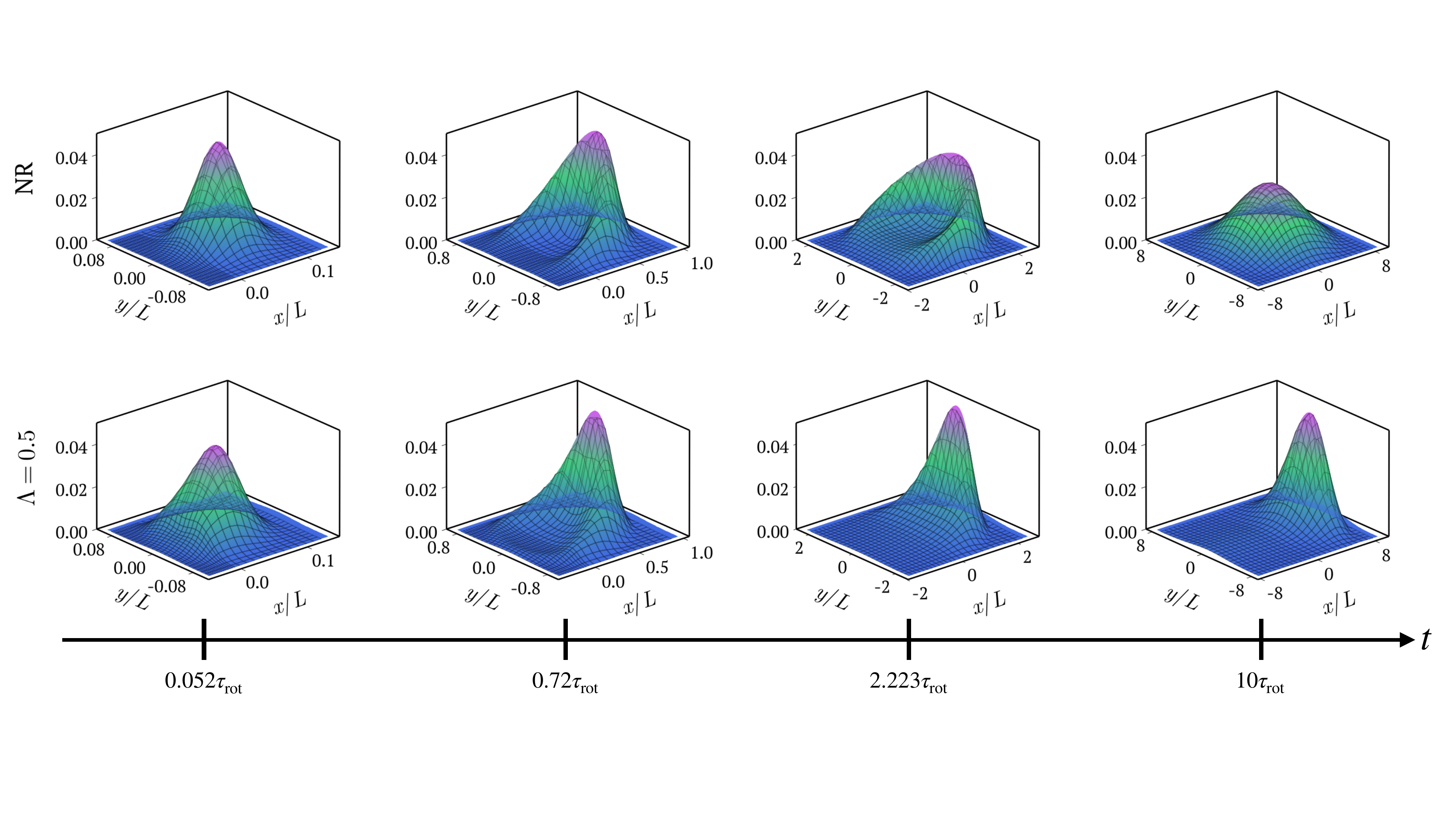}
\caption{Probability distribution function $P(\bm{r},t)$ obtained numerically for a non-resetting ABP with a fixed initial orientation (corresponding to NR) and resetting rate $\Lambda=0.5$. The P\'eclet number is $\mathrm{Pe}=10$, $\tau_{\rm rot}$ denotes the rotational relaxation time, and $L$ is the persistence length.}
     \label{proba_3d}
\end{figure*} 

As a final result, we compute the full probability distribution of the displacements, $P(\bm{r},t)$, by performing an inverse Fourier transform of the ISFs:
\begin{align}
P(\bm{r},t) =  \int_{\mathbb{R}^2} \frac{\mathrm{d}\bm{k}}{(2\pi)^2} \ \widetilde{P}(\bm{k},t) e^{i \bm{k} \cdot \bm{r}}.
\end{align}
Figure~\ref{proba_3d} resolves the full spatial distribution $P(\bm{r},t)$ for several times and gives a broad picture of the agent's dynamics by comparing the non-resetting case with $\Lambda=0.5$. In particular, the transition from $t = 0.052\tau_{\mathrm{rot}}$ to $t = 2.223\tau_{\mathrm{rot}}$ is a good example of how the resetting affects the agent's dynamics. First, we consider the non-resetting case. The probability distribution is asymmetric around the origin, as agents start moving along the $x$-direction. The diametrically-opposed direction does, however, display a non-zero probability as agents reorient due to rotational diffusion. This process leads to a fully symmetric normal distribution at long times, $t = 10.0 \tau_{\mathrm{rot}}$, with a variance that increases as $2D_{\rm eff}^{\mathrm{ ABP}}t$.  

For $\Lambda=0.5$, the probability distribution moves along the resetting direction as time increases. At intermediate times, $t = 0.052\tau_{\mathrm{rot}}$ to $t = 2.223\tau_{\mathrm{rot}}$, it is asymmetric towards the resetting direction and the diametrically-opposed direction displays a much faster decay than the non-resetting counterpart. At times $t = 10.0 \tau_{\mathrm{rot}}$ the distribution of an agent under resetting still exhibits anisotropic features, but we expect it to evolve towards a bivariate Gaussian distribution at even longer times, with mean $\langle\Delta \bm{r}(t)\rangle$ and diagonal covariance matrix $2\mathbf{D}t$ with $\mathbf{D}= \mathrm{diag} [D_{\mathrm{eff}}^{x}, D_{\mathrm{eff}}^{y}]$ (Eqs.~\eqref{eq_mean_disp}, \eqref{deffx}, and~\eqref{deffy}).
This regime is, unfortunately, hard to access numerically due to the long times and large scales that need to be resolved.

\begin{figure*}[tp]
  \centering
  \includegraphics[page=1,width=\textwidth]{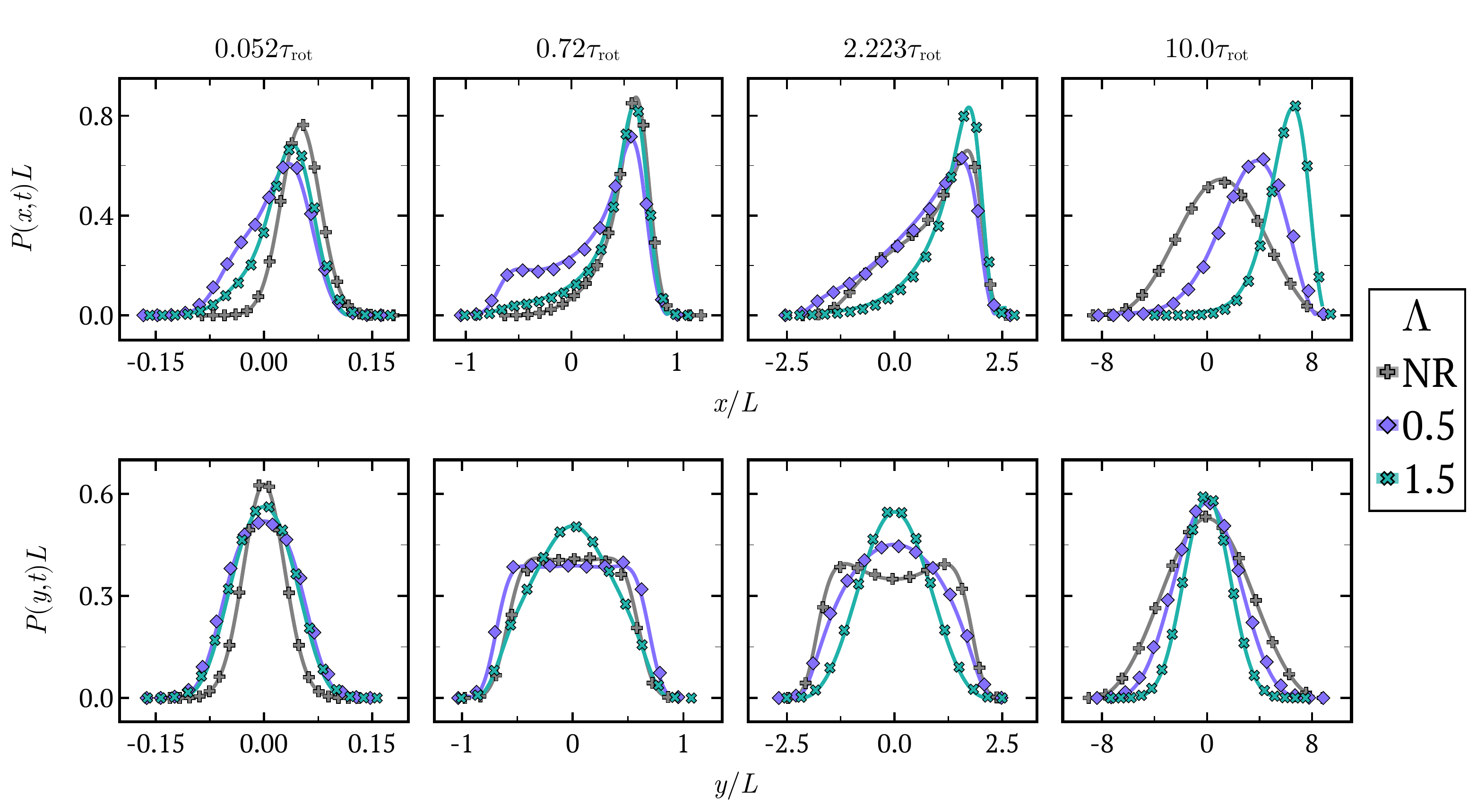}
  \caption{Marginals of the probability distribution function, $P(x,t)$ and $P(y,t)$. The P\'eclet number is $\mathrm{Pe}=10$, $\tau_{\rm rot}$ denotes the rotational relaxation time, and $L$ is the persistence length. Theory and simulation are shown with lines and symbols, respectively.}
     \label{proba_marginals}
\end{figure*} 

To compare different $\Lambda$, we compute the marginals in the resetting and perpendicular directions (e.g. $P(x,t) = \int_{\mathbb{R}}P(\bm{r},t)\mathrm{d}y $), see Fig. \ref{proba_marginals}.  Our semi-analytical results are in excellent agreement with our numerical simulations. In the resetting direction, the distribution becomes increasingly tilted to the right with increasing resetting rate. For $\Lambda=1.5$ the probability to travel along the resetting direction is higher in comparison to a lower resetting rate or to no resetting. In addition, the left tail becomes longer, accounting for the extreme cases where the agent reorients due to rotational diffusion (hence the positive NGP). A notable feature is the shape of the tail for $\Lambda=0.5$ at $t=0.72\tau_{\mathrm{rot}}$ which is due to the non-negligible effect of rotational diffusion. At this $\Lambda$, some particles are still able to reorient, which leads to a larger weight on the left side of the distribution. 

In the $y$-direction, the distribution is always symmetric, as expected from the analysis of the skewness. One can also observe that the width of the distribution decreases with increasing $\Lambda$ at $t\gtrsim 0.72\tau_{\rm rot}$. Due to the effect of resetting, fluctuations around $y=0$ become less and therefore no bimodality appears as in the case of the non-resetting agent. We note that the bimodal distribution of an ABP is reminiscent of the physics of semiflexible polymers. In particular, the probability density for the end-to-end distance transverse to the clamped end of a semiflexible polymer exhibits a bimodal shape \cite{lattanziTransverseFluctuationsGrafted2004,kurzthalerBimodalProbabilityDensity2018}, reflecting the mathematical analogy between ABPs and semiflexible poylmers.

\section{Summary and conclusion}

Here, we have elaborated a renewal framework to characterize the dynamics of an active agent subject to orientational stochastic resetting. We first showed how to use the renewal theory to compute the low-order moments of the displacements and the intermediate scattering function. Not only does the resetting process introduce an overall drift of the agents, it also manifests itself in a non-Gaussian behavior of the probability distributions at intermediate-times (also quantified in terms of a negative skewness and non-zero NGP) followed by an effective diffusion process.

The characterization of such processes in terms of the ISF is of relevance since new alternatives to single-particle tracking have been developed to study the motion of microswimmers~\cite{Cerbino2008, Wilson2011, Martinez2012, Jepson2019, richardsParticleSizingFlowing2021,Bradley2023, Kurzthaler2024, Zhao2024}. Differential dynamic microscopy (DDM) allows to directly measure the ISF and provides statistics in a way that complements single-particle tracking and overcomes some of its shortcomings. Thus, a direct comparison of experimental observations to our theory would provide rapid access to the relevant motility parameters and (chemo)tactic properties of active agents.


The framework of renewal processes can be further expanded to take other mechanisms into account, such as a finite resetting time. In future work, this would be needed when quantitatively describing the resetting process in a complex environment, such as a fluid or a porous medium. The cost of a return to its initial position~\cite{Sunil2023} and the associated energetics~\cite{Fuchs2016, Mori2023} have been computed for a diffusive particle~\cite{Sunil2023}. This aspect is of particular relevance since the introduction of a medium would require the agent to overcome a drag force, which would be associated to a cost and a finite time required to reset the particle's orientation. In addition, it has recently been shown that an ABP exhibits an active contribution to the Taylor dispersion when immersed in a Poiseuille flow~\cite{Peng2020}. Adding a resetting of the reorientation may introduce a coupling to this active contribution and further enhance/suppress the upstream swimming.

Another open question is the determination of the first-passage time properties. The first-passage time distribution has been previously obtained for an ABP exploring a circular region~\cite{Trapani2023}, a generalization for a finite-size target in the resetting case would allow working out the optimal resetting rate to reach this target. Indeed, it has previously been shown that positional stochastic resetting increases the efficiency of foraging~\cite{Reuveni2016, Pal2019, Chechkin2018,Bruyne2020,Przem2022} and even introduces an optimal resetting rate to reach a target, which has been tested experimentally~\cite{Tal2020}. Tackling this question for orientational stochastic resetting would allow for a broader comparison between various search strategies.

Moreover, the first passage time distribution of a one-dimensional Brownian particle subject to spatial stochastic resetting can exhibit a non-monotonic behavior when the resetting position is drawn from a Gaussian distribution \cite{Besga2020,Besga2021}, which would be another incentive to tackle this question given the difficulty of resetting an agent to an exact orientation. This should be done by changing the initial angle distribution $\mathcal{P}(\vartheta)$ that was chosen to be delta distributed in this paper. This theme is of paramount importance when engineering micro-robots where constraints regarding efficiency and biodegradability are imposed.

\begin{acknowledgments}
We thank Nicola Schmidt for discussions at the beginning of the project. T.F. acknowledges funding by FWF: P 35580-N  ``Target Search of Single Active Brownian Particles and Run-and-Tumble Agents”.
\end{acknowledgments}

\onecolumngrid
\section*{Appendix}
\appendix

In Sec.~\ref{ABP_initial_orientation} we present a derivation of the intermediate scattering function of an active Brownian particle (ABP) with a fixed initial orientation and the low-order moments of the displacements. The computation of the stationary distribution of the orientation of a resetting ABP is outlined in Sec.~\ref{evolving_phase_deriv}, while analytical expressions for the moments of the displacements of a resetting ABP are presented in Sec.~\ref{ap:moments_resetting}. Finally, we provide details on numerical procedures, including the computation of the Mathieu functions and the stochastic simulations, in Sec.~\ref{ap:numerics}.

\section{Active Brownian particle with a fixed initial orientation \label{ABP_initial_orientation}}
In this appendix  we present  a derivation of the intermediate scattering function of an active Brownian particle (ABP) with a fixed initial orientation and the low-order moments of the displacements. The solution strategy follows closely the one elaborated in Ref.~\cite{Kurzthaler2018}. 
\subsection{Intermediate scattering function \label{ISF_ABP_NU}}


Here, we follow the solution strategy established in Ref.~\cite{Kurzthaler2018} and extend the results for the case of a fixed initial orientation. Starting from the Langevin equations~\eqref{stoch_r}-\eqref{stoch_theta}, we arrive at the Fokker-Planck equation 
\begin{align}
    \partial_t \mathbb{P}_{\mathrm{ABP}} = -v \bm{u} \cdot \boldsymbol{\nabla}_{\bm{r}} \mathbb{P}_{\mathrm{ABP}} + D\nabla^2_{\bm{r}}\mathbb{P}_{\mathrm{ABP}} + D_{\mathrm{rot}} \partial_\vartheta^2 \mathbb{P}_{\mathrm{ABP}},
\end{align}
\noindent where $\mathbb{P}_{\mathrm{ABP}}(\bm{r}, \vartheta, t | \vartheta_0)$ is the probability density function for the ABP for displacing by $\bm{r}$ in lag time $t$ and displaying an angle $\vartheta$, starting from the initial angle $\vartheta_0$. Performing a spatial Fourier transform $\widetilde{ \mathbb{P}}_{\mathrm{ABP}}(\bm{k}, \vartheta, t | \vartheta_0) = \int_{\mathbb{R}^2}\mathrm{d}^2r \, \mathbb{P}_{\mathrm{ABP}}(\bm{r}, \vartheta, t | \vartheta_0) e^{-i \bm{k} \cdot \bm{r}} $  we obtain:
\begin{align} \label{fpe_fourier}
    \partial_t  \widetilde{ \mathbb{P}}_{\mathrm{ABP}} = -iv \bm{u} \cdot \bm{k} \widetilde{\mathbb{P}}_{\mathrm{ABP}} - k^2 D\widetilde{ \mathbb{P}}_{\mathrm{ABP}} + D_{\mathrm{rot}} \partial_\vartheta^2 \widetilde{ \mathbb{P}}_{\mathrm{ABP}}.
\end{align}
\noindent Without loss of generality, we choose  the wave vector $\bm{k}$ to be $\bm{k} = k \bm{e}_x$. As this aligns $\bm{k}$ with the resetting direction, this choice also simplifies the further expressions. We then have:
\begin{align}
    \partial_t  \widetilde{ \mathbb{P}}_{\mathrm{ABP}} = -ivk \cos(\vartheta)\widetilde{\mathbb{P}}_{\mathrm{ABP}} - k^2 D\widetilde{ \mathbb{P}}_{\mathrm{ABP}} + D_{\mathrm{rot}} \partial_\vartheta^2 \widetilde{ \mathbb{P}}_{\mathrm{ABP}}. \label{eq:fourier}
\end{align}
\noindent As suggested by the nature of the equation, we try to find a solution by separating the variables: $e^{-\lambda t} z(x)$ where $x = \vartheta/2$. This leads to the eigenvalue problem for the eigenfunctions $z(x)$:
\begin{align}
    \left[ \frac{\mathrm{d}^2}{\mathrm{d}x^2} - \frac{4ivk}{D_{\mathrm{rot}}} \cos(2x)\right] z(x) = \frac{4}{D_{\mathrm{rot}}}(Dk^2-\lambda) z(x).
\end{align}
\noindent Introducing the complex parameter $q = 2ivk/D_{\mathrm{rot}}$ and $a= 4(\lambda - Dk^2)/D_{\mathrm{rot}}$, we obtain the Mathieu equation~\cite{NIST:DLMF}:
\begin{align}
    \left[\frac{\mathrm{d}^2}{\mathrm{d}x^2} + \Bigl( a -2q\cos(2x)\Bigl) \right] z(x) = 0. \label{eq:mathieu}
\end{align}
As $z$ is $\pi$-periodic, the solutions are the even and odd Mathieu functions: 
\begin{subequations}
\begin{align}
    \operatorname{ce}_{2n}(x,q) &= \sum_{p=0}^{+\infty} A_{2p}^{2n}(q) \cos(2px), \label{eq:mathieu_even}\\ 
    \operatorname{se}_{2n+2}(x,q) &= \sum_{p=0}^{+\infty} B_{2p+2}^{2n+2}(q) \sin \Bigl( (2p+2)x \Bigl), \label{eq:mathieu_odd}
\end{align}
\end{subequations}
\noindent with associated eigenvalues $a_{2n}(q)$ and $b_{2n+2}(q)$ for $n \in \mathbb{N}$. The Fourier coefficients $A_{2p}^{2n}(q)$, upon inserting Eq.~\eqref{eq:mathieu_even} into the Mathieu equation [Eq.~\eqref{eq:mathieu}], can be shown to follow the recursion relations:
\begin{subequations}
\begin{align}
    a_{2n} A_0^{2n} &= q A_2^{2n}, \\
    (a_{2n}-4) A_2^{2n} &= q (2A_0^{2n}+A_4^{2n}), \\
    (a_{2n}-4p^2) A_{2p}^{2n}p^{2n} &= q (A_{2p-2}^{2n}+A_{2p+2}^{2n}).
    \label{eigenval_rec}
\end{align}
\end{subequations}
Similar relations hold for $B_{2p+2}^{2n+2}(q)$, see Ref.~\cite{NIST:DLMF}. 
To express the full solution of Eq.~\eqref{eq:fourier} as a superposition of the Mathieu functions, we first show that they are orthogonal and normalized in the following sense: 
\begin{subequations}
\begin{align}
    \frac{1}{\pi} \int_0^{2\pi} \mathrm{d}x \operatorname{ce}_{2n}(x,q) \operatorname{ce}_{2m}(x,q)  &=  \delta_{nm},\label{eq:scalar}\\
    \frac{1}{\pi} \int_0^{2\pi} \mathrm{d}x \operatorname{se}_{2n+2}(x,q) \operatorname{se}_{2m+2}(x,q)  &=  \delta_{nm}.
\end{align}
\end{subequations}
For two $2\pi$-periodic integrable functions, $\phi$ and $\psi$, we define the scalar product by 
\begin{equation}
    \langle \phi | \psi \rangle = \frac{1}{\pi} \int_0^{2\pi} \mathrm{d}x \ \overline{\phi(x)} \psi(x),
\end{equation}
where $\overline{\phi(x)}$ is the complex conjugate of $\phi(x)$. Then the Mathieu functions are orthonormalized with respect to this scalar product. We then introduce the non-Hermitian Sturm-Liouville operator: 
\begin{equation}
    \widehat{\mathcal{H}}(q) = - \frac{\mathrm{d}^2}{\mathrm{d}x^2} + 2q \cos(2x),
\end{equation}
with right eigenfunction $R_m(x,q)$ and eigenvalue $r_m(x,q)$ such that $\widehat{\mathcal{H}} R_m = r_m R_m$. Then its  adjoint operator obeys
\begin{equation}
    \widehat{\mathcal{H}}^\dagger(q) = - \frac{\mathrm{d}^2}{\mathrm{d}x^2} + 2 \overline{q} \cos(2x),
\end{equation}
with left eigenfunction $L_m(x,q)$ and eigenvalue $l_m(x,q)$ such that $\widehat{\mathcal{H}}^\dagger L_m = \overline{l_m} L_m$.
The following equality then holds:
\begin{equation}
    \langle L_m | \widehat{\mathcal{H}} R_n \rangle = r_n \langle L_m | R_n \rangle = \langle \widehat{\mathcal{H}}^\dagger L_m  | R_n \rangle = l_m \langle L_m | R_n \rangle.
\end{equation}
Thus, it  follows that:
\begin{equation}
    (r_n -l_m) \langle L_m | R_n \rangle = 0,
\end{equation}
and, consequently, the eigenfunctions corresponding to different eigenvalues are orthogonal. We further make the assumption that the spectrum is non-degenerate, and in the case that the eigenvalues $r_n$ and $l_m$ are the same, we label right and left eigenfunctions to eigenvalue $r_m$ by $\langle L_m|$ and $| R_m \rangle $, respectively. We then compute:
\begin{align}
    \overline{\widehat{\mathcal{H}}^\dagger(q) L_n(x,q)} &=  \Bigl[ - \frac{\mathrm{d}^2}{\mathrm{d}x^2} + 2q \cos(2x)  \Bigr] \overline{L_n(x,q)} 
    =  l_n(q) \overline{L_n(x,q)} = r_n(q) \overline{L_n(x,q)} 
    = \widehat{\mathcal{H}} \overline{L_n(x,q)}.
\end{align}
We observe that up to normalization, $\overline{L_n(x,q)} \propto R_n(x,q)$. We choose them to be equal and normalize them such that $\langle L_n | R_m \rangle = \delta_{n,m}$. Thus, using Eq.~\eqref{eq:scalar} we find that the Fourier coefficients obey \cite{Ziener2012, NIST:DLMF}:
\begin{equation}
    2[A_0^{2n}(q)]^2 + \sum_{p=1}^{+\infty} [A_{2p}^{2n}(q)]^2 = 1.
\end{equation}

Finally, the solution to Eq.~\eqref{eq:fourier} can be written as a superposition of the appropriate eigenfunctions:

\begin{equation}
\begin{split}
    \widetilde{\mathbb{P}}_{\mathrm{ABP}}(\bm{k},x,t | x_0) = e^{-k^2 D t} \sum_{n=0}^{+\infty} \left( \mu_{2n}^a \operatorname{ce}_{2n} \left( x,q \right) e^{-D_{\mathrm{rot}} a_{2n}(q)t/4} + \mu_{2n+2}^b \operatorname{se}_{2n+2} \left( x,q \right) e^{-D_{\mathrm{rot}} b_{2n+2}(q) t/4} \right) .
\end{split}
\end{equation}
\noindent Imposing the initial condition $\widetilde{\mathbb{P}}(\bm{k},x, 0 | x_0) = \frac{1}{2}\delta(x-x_0)$ and using that the Mathieu functions constitute a complete, orthogonal, and normalized set of eigenfunctions, we arrive at the following expression for the coefficients:
\begin{subequations}
\begin{align}
    \mu_{2n}^a &= \frac{1}{\pi} \int_0^{2\pi}\mathrm{d}x \frac{1}{2}\delta(x-x_0) \operatorname{ce}_{2n}(x,q) = \frac{1}{2\pi} \operatorname{ce}_{2n}(x_0,q), \\
    \mu_{2n+2}^b &= \frac{1}{\pi} \int_0^{2\pi}\mathrm{d}x \frac{1}{2}\delta(x-x_0) \operatorname{se}_{2n+2}(x,q) = \frac{1}{2\pi} \operatorname{se}_{2n+2}(x_0,q).
\end{align}
\end{subequations}

\noindent Going back to the variable $\vartheta$, the solution is:
\begin{equation}
\begin{split}
    \widetilde{\mathbb{P}}_{\mathrm{ABP}}(\bm{k},\vartheta,t | \vartheta_0) = \frac{1}{2\pi} e^{-k^2 D t} \sum_{n=0}^{+\infty}\Bigg( &\operatorname{ce}_{2n} \left( \frac{\vartheta_0}{2},q \right) \operatorname{ce}_{2n} \left( \frac{\vartheta}{2},q \right) e^{-D_{\mathrm{rot}}a_{2n}(q) t/4}\\ 
    &+ \operatorname{se}_{2n+2} \left( \frac{\vartheta_0}{2},q \right) \operatorname{se}_{2n+2} \left( \frac{\vartheta}{2},q \right) e^{-D_{\mathrm{rot}}b_{2n+2}(q)t/4} \Bigg) .
\end{split}
\end{equation}

\noindent We obtain the ISF by marginalizing over the angle $\vartheta$ and the initial angle $\vartheta_0$, taking into account that it is initially fixed along the resetting direction (see Eq.~\eqref{eq:p} in the main text). We then compute $\widetilde{\mathbb{P}}(\bm{k},t)$ via 
\begin{align}
    \widetilde{\mathbb{P}}(\bm{k},t) = \int_0^{2\pi} \mathrm{d}\vartheta \int_0^{2\pi} \mathrm{d}\vartheta_0 \, \widetilde{\mathbb{P}}_{\mathrm{ABP}}(\bm{k},\vartheta,t | \vartheta_0)\mathcal{P}(\vartheta_0)
    = e^{-k^2 Dt} \sum_{n=0}^{+\infty} A_{0}^{2n}(q)\operatorname{ce}_{2n} \left(0,q \right) e^{-D_{\mathrm{rot}}a_{2n}(q)t/4},
    \label{ISF_free}
\end{align}
where we used $\mathcal{P}(\vartheta_0)=\delta(\vartheta_0)$.
\subsection{Low-order moments of the displacements \label{appendix_abp_moments}}
Based on \cite{Hagen2011, hagenNonGaussianBehaviourSelfpropelled2009}, we provide low-order moments of an ABP with initial orientation $\vartheta_{0} =0$. Defining $\langle \cdot \rangle_{\mathrm{ABP}}$ as the average over noise realizations for a simple ABP starting with initial orientation $\vartheta_0=0$ , the mean displacements evaluate to 
\begin{subequations}
\begin{align}
    \langle \Delta x(t) \rangle_{\mathrm{ABP}} &= \frac{v}{D_{\mathrm{rot}}} (1-e^{-D_{\mathrm{rot}} t}),\\
    \langle \Delta y(t) \rangle_{\mathrm{ABP}} &= 0.
\end{align}
\end{subequations}
The second-order moments read:
\begin{subequations}
\begin{align}
    \langle \Delta x(t)^2 \rangle_{\mathrm{ABP}} &= 2Dt + \frac{v^2}{D_{\mathrm{rot}}^2} \biggr[ D_{\mathrm{rot}} t +e^{-D_{\mathrm{rot}} t} - 1 + \frac{1}{12} (3 + e^{-4 D_{\mathrm{rot}} t} - 4e^{-D_{\mathrm{rot}} t})\biggr],\\
    \langle \Delta y(t)^2 \rangle_{\mathrm{ABP}} &= 2Dt + \frac{v^2}{D_{\mathrm{rot}}^2} \biggr[ D_{\mathrm{rot}} t +e^{-D_{\mathrm{rot}} t} - 1 - \frac{1}{12} (3 + e^{-4 D_{\mathrm{rot}} t} - 4e^{-D_{\mathrm{rot}} t})\biggr].
\end{align}
\end{subequations}
The third-order moments are
\begin{subequations}
\begin{align}
    \begin{split}
    \langle \Delta x(t)^3 \rangle_{\mathrm{ABP}} &= \frac{6Dv}{D_{\mathrm{rot}}^2}(1-e^{-D_{\mathrm{rot}} t})D_{\mathrm{rot}} t +  \frac{v^3}{D_{\mathrm{rot}}^3} \biggr[ -\frac{45}{8} + \frac{1}{24} + \frac{17}{3}e^{-D_{\mathrm{rot}} t} -\frac{1}{16}e^{-D_{\mathrm{rot}} t}  \\ 
    &\quad + 3D_{\mathrm{rot}} t - \frac{1}{24}e^{-4D_{\mathrm{rot}} t} + \frac{1}{40} e^{-4D_{\mathrm{rot}} t}  + \frac{5}{2}e^{-D_{\mathrm{rot}} t}D_{\mathrm{rot}} t  -\frac{1}{240}e^{-9D_{\mathrm{rot}} t} \biggr],
    \end{split}\\
    \begin{split}
    \langle \Delta y(t)^3 \rangle_{\mathrm{ABP}} &= 0,
    \end{split}
\end{align}
\end{subequations}
and the fourth-order moments assume the form
\begin{subequations}
\begin{align}
    \begin{split}
    \langle \Delta x(t)^4 \rangle_{\mathrm{ABP}} &= 12D^2 t^2 + 12Dt \Bigl(\frac{v}{D_{\mathrm{rot}}}\Bigl)^2 \biggr[ e^{-D_{\mathrm{rot}} t} + D_{\mathrm{rot}} t -1 + \frac{1}{12} \Bigl( e^{-4D_{\mathrm{rot}} t} -4e^{-D_{\mathrm{rot}} t} +3  \Bigl) \biggr]  \\ 
    &\quad+   \Bigl(\frac{v}{D_{\mathrm{rot}}}\Bigl)^4 \biggr[ 3(D_{\mathrm{rot}} t)^2 + \frac{1}{6720}e^{-16 D_{\mathrm{rot}} t} -5D_{\mathrm{rot}} t e^{-D_{\mathrm{rot}} t}  - \frac{45}{4} D_{\mathrm{rot}} t + \frac{261}{16} + \frac{1}{600} e^{-9 D_{\mathrm{rot}} t}\\ 
    &\quad-\frac{19}{6}  + \frac{1}{240}e^{-4D_{\mathrm{rot}} t} + \frac{1}{192} + \frac{1}{48} e^{-4D_{\mathrm{rot}} t} - \frac{49}{3} e^{-D_{\mathrm{rot}} t} - \frac{7}{450}e^{-4D_{\mathrm{rot}} t} - \frac{1}{120}e^{-D_{\mathrm{rot}} t} \\
    &\quad+ \frac{3}{2}D_{\mathrm{rot}} t - \frac{1}{30}e^{-4D_{\mathrm{rot}} t}D_{\mathrm{rot}} t + \frac{229}{72} e^{-D_{\mathrm{rot}} t} - \frac{1}{840} e^{-9 D_{\mathrm{rot}} t} + \frac{5}{3}e^{-D_{\mathrm{rot}} t} D_{\mathrm{rot}} t \biggr],
    \end{split}\\
    \begin{split}
    \langle \Delta y(t)^4 \rangle_{\mathrm{ABP}} &= 12D^2 t^2 + 12Dt \Bigl(\frac{v}{D_{\mathrm{rot}}}\Bigl)^2 \biggr[ e^{-D_{\mathrm{rot}} t} + D_{\mathrm{rot}} t -1 - \frac{1}{12} \Bigl( e^{-4D_{\mathrm{rot}} t} -4e^{-D_{\mathrm{rot}} t} +3  \Bigl) \biggr]  \\ 
    &\quad+   \Bigl(\frac{v}{D_{\mathrm{rot}}}\Bigl)^4 \biggr[ 3(D_{\mathrm{rot}} t)^2 + \frac{1}{6720}e^{-16 D_{\mathrm{rot}} t} -5D_{\mathrm{rot}} t e^{-D_{\mathrm{rot}} t}  - \frac{45}{4} D_{\mathrm{rot}} t + \frac{261}{16} - \frac{1}{600} e^{-9 D_{\mathrm{rot}} t}\\ 
    &\quad+ \frac{19}{6}  + \frac{1}{240}e^{-4D_{\mathrm{rot}} t} + \frac{1}{192} + \frac{1}{48} e^{-4D_{\mathrm{rot}} t} - \frac{49}{3} e^{-D_{\mathrm{rot}} t} + \frac{7}{450}e^{-4D_{\mathrm{rot}} t} - \frac{1}{120}e^{-D_{\mathrm{rot}} t} \\
    &\quad- \frac{3}{2}D_{\mathrm{rot}} t + \frac{1}{30}e^{-4D_{\mathrm{rot}} t}D_{\mathrm{rot}} t - \frac{229}{72} e^{-D_{\mathrm{rot}} t} - \frac{1}{840} e^{-9 D_{\mathrm{rot}} t} - \frac{5}{3}e^{-D_{\mathrm{rot}} t} D_{\mathrm{rot}} t \biggr].
    \end{split}
\end{align}
\end{subequations}

\section{Stationary distribution of the orientation of a resetting agent\label{evolving_phase_deriv}}

We provide a derivation for the computation of the stationary distribution of the angle. The strategy consists in upgrading the renewal equations by explicitly keeping the dependence on the orientation. Therefore, we introduce the quantities
\begin{subequations}
\begin{align}
\mathbb{P}^0(\bm{r}, \vartheta, t) &= \int_0^{2\pi} \mathrm{d}\vartheta_0 \mathbb{P}_{\mathrm{ABP}}(\bm{r}, \vartheta, t | \vartheta_0)p^0(\vartheta_0), \\
\mathbb{P}(\bm{r}, \vartheta, t) &= \int_0^{2\pi} \mathrm{d}\vartheta_0 \mathbb{P}_{\mathrm{ABP}}(\bm{r}, \vartheta, t | \vartheta_0) \mathcal{P}(\vartheta_0),
\end{align}
\end{subequations}
and rewrite the renewal equations (in Fourier space) by keeping the angular dependence:
\begin{subequations}
\begin{align}
	\widetilde{Q}(\bm{k}, t) &= \widetilde{Q}_1(\bm{k}, t) + \int_0^t\mathrm{d}t' \widetilde{Q}(\bm{k},t-t') T(t')\widetilde{\mathbb{P}}(\bm{k}, t'), \label{ap:qtilde} \\
	\widetilde{P}(\bm{k}, \vartheta, t) &= \widetilde{P}_0(\bm{k}, \vartheta, t)+ \int_0^t\mathrm{d}t' \widetilde{Q}(\bm{k},t-t') S(t') \widetilde{\mathbb{P}}(\bm{k}, \vartheta, t') , \label{eq_p_theo} \\
	\widetilde{Q}_1(\bm{k}, t) &= \widetilde{\mathbb{P}}^0(\bm{k},t) \frac{1}{\tau}S(t), \label{ap:q0tilde}\\ 
	\widetilde{P}_0(\bm{k},\vartheta, t) &= \widetilde{\mathbb{P}}^0(\bm{k},\vartheta,t) \frac{1}{\tau}\int_t^{\infty}\mathrm{d}t' S(t').
\end{align}
\end{subequations}
As the change of orientation is not coupled to the position in the Fokker-Planck equation, we can obtain the distribution of $\vartheta$ by setting $\bm{k}=\bm{0}$: $P(\vartheta, t) = \widetilde{P}(\bm{k}= \bm{0}, \vartheta,t)$ (and, similarly, $P_0(\vartheta, t) = \widetilde{P}_0(\bm{k}= \bm{0}, \vartheta,t)$ and $\mathbb{P}(\vartheta,t) = \widetilde{\mathbb{P}}(\bm{k} = \bm{0}, \vartheta, t)$). The stationary distribution of the angle is then obtained via
\begin{equation}
p^0(\vartheta) = \underset{t \to \infty}{\lim} P(\vartheta,t).
\end{equation}
We note that due to normalization we have $\widetilde{\mathbb{P}}(\textbf{k}=\textbf{0},t)=1$ and $\widetilde{\mathbb{P}}^0(\textbf{k}=\textbf{0},t) = 1$ and that at long times the probability to never have reset vanishes, $\underset{t \to \infty}{\lim} P_0(\vartheta, t) = 0$. Performing a Laplace transform of Eqs.~\eqref{ap:qtilde}-\eqref{ap:q0tilde}, defining $\widehat{Q}(s) = \mathcal{L}[\widetilde{Q}(\bm{k}=\bm{0},t)](s) $, and  $\widehat{Q}_1(s) = \mathcal{L}[\widetilde{Q}_1(\bm{k}=\bm{0},t)](s) $, we arrive at:

\begin{equation}
    \widehat{Q}(s) = \frac{1}{\tau} \frac{\widehat{S}(s)}{1-\widehat{T}(s)} = \frac{1}{s \tau},
\end{equation}
where we have used $\widehat{S}(s) = (1-\widehat{T}(s))/s$. The previous equation further implies 

\begin{equation} \label{q_statio}
    Q(t) = \frac{1}{\tau} = \mathrm{const.} \in \mathbb{R},
\end{equation}
demonstrating that the renewal events are in a stationary state and occur with time-independent rate $1/\tau$.
Taking the limit $t \to \infty$ for $\bm{k}=\bm{0}$ in Eq.~\eqref{eq_p_theo} and using Eq.~\eqref{q_statio} we arrive at
\begin{equation} \label{statio_proba}
p^0(\vartheta)  = \underset{t\to \infty}{\lim} P(\vartheta, t) = \frac{1}{\tau} \int_{0}^{\infty}\mathrm{d}t'S(t') \mathbb{P}(\vartheta, t').
\end{equation}
This result holds for arbitrary waiting time distributions with finite mean $\tau$ and generalizes the case of exponential waiting times studied in the physics literature \cite{Evans2020}.
There are several ways to compute the distribution of the angle $\mathbb{P}(\vartheta, t)$. One approach relies on a separation of variables and expanding $\mathbb{P}(\vartheta, t)$ in terms of the angular eigenfunctions $e^{i n\vartheta}$ with $n\in\mathbb{Z}$. Another approach is to compute it by first ignoring the periodicity of the angle. In this case, the problem can be mapped to that of resetting the position of a Brownian particle to a fixed position at exponentially distributed times, as discussed in the main text [Eq.~\eqref{eq:stationary_BP}]~\cite{Evans2011}. Since in our case, the orientation is reset, instead of the particle's position, we need to wrap it around $[-\pi$, $\pi]$. If we let $\mathbb{P}$ be the probability associated to the random variable $\Theta_t$ of density $p^{0}_{\mathrm{pos}}$, and $A \subseteq [-\pi, \pi]$, we have:
\begin{subequations}
\begin{align}
    \mathbb{P}(\Theta_t \mod 2\pi \in A)  &= \mathbb{P}(\Theta_t \in A + 2\pi n \ \text{for some $n \in \mathbb{N}$}), \\
    &= \sum_{n=-\infty}^{+\infty} \mathbb{P}(\Theta_t \in A + 2\pi n), \\ 
    &= \sum_{n=-\infty}^{+\infty} \int_{A + 2\pi n} \mathrm{d} \vartheta p^0_{\mathrm{pos}}(\vartheta) , \\ 
    &= \int_A \mathrm{d} \vartheta \underbrace{\sum_{n=-\infty}^{+\infty} p^0_{\mathrm{pos}}(\vartheta + 2\pi n)}_{\text{density} \  p^0(\vartheta)} \ \text{by normal convergence}.
\end{align}
\end{subequations}
\noindent Finally, the sum reduces to
\begin{align}
    p^{0}(\vartheta)&= \sum_{m={-\infty}}^{+\infty} \frac{\sqrt{\Lambda}}{2}  e^{-\sqrt{\Lambda}|\vartheta + 2\pi m|} = \frac{\sqrt{\Lambda}}{2} \frac{\cosh(\sqrt{\Lambda}(\pi-|\vartheta|))}{\sinh(\sqrt{\Lambda}\pi)}, 
\end{align}
which is the result presented in the main text, Eq. \eqref{closed_p0}.

\section{Low-order moments of the displacements \label{ap:moments_resetting}}

Here, we present the moments of the displacements for the ABP under orientational stochastic resetting with $T(t)=\lambda e^{-\lambda t}$ and $\mathcal{P}(\vartheta)=\delta(\vartheta)$. The mean displacements are:
\begin{subequations}
\begin{align}
    \langle \Delta x(t) \rangle &= \frac{\lambda v}{D_\mathrm{rot}+ \lambda}t,\\
    \langle \Delta y(t) \rangle &= 0.
\end{align}
\end{subequations}
The second-order moments, from which the variances [Eqs.~\eqref{vx}-\eqref{vy}] were derived, are the following:
\begin{subequations}
\begin{align}
\begin{split}
\langle \Delta x(t)^2 \rangle &= 2Dt - \frac{2D_{\mathrm{rot}}^{2} v^{2} ( 1- e^{-(D_{\mathrm{rot}}+\lambda)t} )(2D_{\mathrm{rot}}+5\lambda) }{(D_{\mathrm{rot}}+\lambda)^{4} (4D_{\mathrm{rot}}+ \lambda)}\\ &\quad + \frac{tv^{2} \Big( 2D_{\mathrm{rot}}^{2}(2D_{\mathrm{rot}}+5 \lambda) + \lambda^{2}(D_{\mathrm{rot}}+\lambda)(4D_{\mathrm{rot}}+\lambda)t \Big)}{(D_{\mathrm{rot}}+\lambda)^{3}(4D_{\mathrm{rot}}+ \lambda)},
\end{split}\\
\begin{split}
\langle \Delta y(t)^2 \rangle &= 2Dt + \frac{4D_{\mathrm{rot}} v^{2} \Big( e^{-(D_{\mathrm{rot}}+\lambda)t} -1 + (D_{\mathrm{rot}}+\lambda)t \Big)}{(D_{\mathrm{rot}}+\lambda)^{2} (4D_{\mathrm{rot}}+ \lambda)}. 
\end{split}
\end{align}
\end{subequations}

The third-order moments read:
\begin{subequations}
\begin{align}
\begin{split}
\langle \Delta x(t)^3 \rangle &= \frac{  \lambda v}{(D_{\mathrm{rot}} +\lambda) (4D_{\mathrm{rot}} +\lambda) (9D_{\mathrm{rot}} +\lambda)} \Bigg( -\frac{20D_{\mathrm{rot}}e^{-(4D_{\mathrm{rot}}+\lambda)t}v^2 }{(4D_{\mathrm{rot}}+\lambda)^2} \\ 
&\quad- \frac{4D_{\mathrm{rot}} e^{-(D_{\mathrm{rot}}+\lambda)t}v^2 (103 D_{\mathrm{rot}}^{3} + 267 D_{\mathrm{rot}}^{2} \lambda + 15D_{\mathrm{rot}} \lambda^{2} -5\lambda^{3}) }{(D_{\mathrm{rot}}+\lambda)^5} \\
&\quad+ \frac{12 D_{\mathrm{rot}}^{3}v^2 (551 D_{\mathrm{rot}}^{3} + 1707 D_{\mathrm{rot}}^{2} \lambda + 843 D_{\mathrm{rot}} \lambda^{2} + 119 \lambda^{3}) }{(D_{\mathrm{rot}}+\lambda)^{5}(4D_{\mathrm{rot}}+\lambda)^{2}} \\ 
&\quad- \frac{6D_{\mathrm{rot}}^{2} e^{-(D_{\mathrm{rot}}+\lambda)t}tv^{2} (9D_{\mathrm{rot}}+\lambda)(2D_{\mathrm{rot}}+5\lambda) }{(D_{\mathrm{rot}}+\lambda)^{4}}  \\
&\quad- \frac{6D_{\mathrm{rot}}^{2}tv^{2} (206D_{\mathrm{rot}}^{3} +588 D_{\mathrm{rot}}^{2}
\lambda +171 D_{\mathrm{rot}} \lambda^{2} + 5\lambda^{3}) }{(D_{\mathrm{rot}}+\lambda)^{4}(4D_{\mathrm{rot}}+\lambda)} 
+ \frac{\lambda^{2} (4D_{\mathrm{rot}}+\lambda)(9D_{\mathrm{rot}}+\lambda) t^{3}v^{2}}{(D_{\mathrm{rot}}+\lambda)^2} \\
&\quad+ \frac{6t^{2}(9D_{\mathrm{rot}}+\lambda) \big(D(D_{\mathrm{rot}}+\lambda)^{3}(4D_{\mathrm{rot}} +\lambda) + D_{\mathrm{rot}}^{2}(2D_{\mathrm{rot}}+5\lambda) v^2 \big)}{(D_{\mathrm{rot}}+\lambda)^{3}}
\Bigg),
\end{split}\\
\begin{split}
\langle \Delta y(t)^3 \rangle &= 0. 
\end{split}
\end{align}
\end{subequations}

We conclude with the fourth-order moments:
\begin{subequations}
\begin{align}
\langle \Delta x(t)^{4} \rangle &= 12D^{2}t^{2} + \frac{12De^{-(D_{\mathrm{rot}}+\lambda)t} tv^{2}}{(D_{\mathrm{rot}}+\lambda)^4 (4D_{\mathrm{rot}} +\lambda)}\Bigg( 2D_{\mathrm{rot}}^{2} (2D_{\mathrm{rot}}+5\lambda) \notag\\ 
&\quad+ e^{(D_{\mathrm{rot}}+\lambda)t} \Big( 4D_{\mathrm{rot}}^{4}t + 6D_{\mathrm{rot}}t^{2} \lambda^{4} + t^{2} \lambda^{5} + 2D_{\mathrm{rot}}^{3} \big(-2 + t\lambda (7+2t \lambda) \big) + D_{\mathrm{rot}}^{2} \lambda \big(-10 + t\lambda (10+9t\lambda ) \big) \Big) \Bigg) \notag\\
&\quad+ \frac{e^{-14D_{\mathrm{rot}}t -3\lambda t} v^4 }{12(D_{\mathrm{rot}}+\lambda)^{8} (4D_{\mathrm{rot}}+ \lambda)^{4} (9D_{\mathrm{rot}}+ \lambda)^3 (16 D_{\mathrm{rot}} + \lambda)} \Bigg( \notag\\
&\quad107495424D_{\mathrm{rot}}^{14} e^{14D_{\mathrm{rot}}t + 3\lambda t}t^2 + 3e^{5D_{\mathrm{rot}}t + 2\lambda t} \lambda^{12}(27-192e^{5D_{\mathrm{rot}}t} +165e^{8D_{\mathrm{rot}}t} +4e^{(9D_{\mathrm{rot}}+\lambda)t}t^4 \lambda^4) \notag\\
&\quad+ D_{\mathrm{rot}}e^{5D_{\mathrm{rot}}t + 2\lambda t} \lambda^{11} \bigg(1701 -20672e^{5D_{\mathrm{rot}}t} + 756e^{(9D_{\mathrm{rot}}+\lambda)t}t^{4} \lambda^{4} + e^{8D_{\mathrm{rot}}t}(18971 - 1080 \lambda t) \bigg)  \notag\\
&\quad+ 2239488D_{\mathrm{rot}}^{13} e^{13D_{\mathrm{rot}}t + 2\lambda t}t \bigg( -80 +e^{(D_{\mathrm{rot}}+\lambda)t} \big( -180 + \lambda t (379 +96 \lambda t) \big) \bigg) \notag\\
&\quad+D_{\mathrm{rot}}^{2} e^{5D_{\mathrm{rot}}t + 2\lambda t} \lambda^{10} \bigg( 15471 -276736 e^{5D_{\mathrm{rot}}t} + e^{8D_{\mathrm{rot}}t} \big( 261265 + 24\lambda t(-2017+30 \lambda t) \big) \notag\\ 
&\quad+36e^{(9D_{\mathrm{rot}}+\lambda)t} t^{2} \lambda^{2} \big( -20 + t\lambda (20 +567 t\lambda) \big) \bigg) \notag\\ 
&\quad+D_{\mathrm{rot}}^{3}e^{5D_{\mathrm{rot}}t + 2\lambda t} \lambda^{9} \bigg( 80433 -1700480 e^{5D_{\mathrm{rot}}t} +e^{8D_{\mathrm{rot}}t} \big( 1620047 +168 t \lambda (-4613+246 t \lambda)  \big) \notag\\
&\quad+ 36 e^{(9D_{\mathrm{rot}} + \lambda)t} \lambda t\Big( 1904 + t \lambda \big[-1980 + t \lambda (1168 + 8665 t\lambda ) \big] \Big) \bigg)  \notag\\ 
&\quad+ 124416D_{\mathrm{rot}}^{12} e^{10D_{\mathrm{rot}}t + 2\lambda t} \bigg( 6 + 6e^{3D_{\mathrm{rot}}t} \big( -784 + t \lambda (-1643 +288 t\lambda) \big) \notag\\  
&\quad+ e^{(4D_{\mathrm{rot}}+\lambda)t} \Big( 4698 + t\lambda \big[ -22335 + t\lambda \big( 12437 +36 t \lambda (319 +8t \lambda) \big) \big] \Big) \bigg) \notag\\
&\quad+ 576 D_{\mathrm{rot}}^{11}e^{5D_{\mathrm{rot}}t + 2\lambda t} \lambda \bigg( 18 \Big( 1+609e^{5D_{\mathrm{rot}}t} + e^{8D_{\mathrm{rot}}t} \big( -332385 + 2t \lambda (-61499 +58536 t \lambda) \big)  \Big)\notag\\ 
&\quad+ e^{(9D_{\mathrm{rot}}+\lambda)t} \Big( 5 971 950 + \lambda t \big[ -6 161 166 + t\lambda \big(-3 360 371 + 216\lambda t (29971 + 1554 t\lambda) \big) \big]  \Big)  \bigg)  \notag\\
&\quad+ D_{\mathrm{rot}}^{4}e^{5D_{\mathrm{rot}}t + 2\lambda t} \lambda^{8} \bigg( 266 085 -4 681 600 e^{5D_{\mathrm{rot}}t} + e^{8D_{\mathrm{rot}}t} \Big( 9 033 595 + 24 t \lambda \big( -177 289 +41 184 t\lambda \big) \Big)  \notag\\
&\quad+ 36e^{(9D_{\mathrm{rot}}+\lambda)t} \Big( -128 280 + t\lambda \big( 122 640 + t\lambda \big[-67 020 + t \lambda (28 604 + 83 265 t \lambda) \big] \big) \Big) \bigg) \notag\\
&\quad+ 96D_{\mathrm{rot}}^{10}e^{5D_{\mathrm{rot}}t + 2\lambda t} \lambda^2 \bigg( 891 +239 616 e^{5D_{\mathrm{rot}}t} + e^{8D_{\mathrm{rot}}t} \Big( -15 385 563 +2t \lambda \big( 24 627 787 + 13 099 320 t\lambda \big) \Big)  \notag\\
&\quad+ 12 e^{(9D_{\mathrm{rot}}+\lambda)t} \Big(1 262 088 + t\lambda \big( 6 763 830 + t\lambda \big[ -7 585 950 + t\lambda (4 413 757 + 386 064 t\lambda) \big] \big) \Big) \bigg)  \notag\\
&\quad+ D_{\mathrm{rot}}^{5} e^{5D_{\mathrm{rot}}t + 2\lambda t} \lambda^{7} \bigg( 590 247 -3 582 080 e^{5D_{\mathrm{rot}}t} + e^{8D_{\mathrm{rot}}t} \Big( 125 070 713 + 24t \lambda \big( 897 919 + 537 540 t\lambda \big) \Big) \notag\\ 
&\quad+ 36e^{(9D_{\mathrm{rot}}+\lambda)t} \Big( -3 391 080 + t\lambda \big( 2 589 744 + t\lambda \big[-1 105 860 + t\lambda (385 816 + 528 255 t\lambda) \big] \big) \Big) \bigg) \notag\\ 
&\quad+D_{\mathrm{rot}}^{6} e^{5D_{\mathrm{rot}}t + 2\lambda t} \lambda^{6} \bigg(899 829 + 12 266 240e^{5D_{\mathrm{rot}}t} + e^{8D_{\mathrm{rot}}t} \Big( 1 232 304 619 + 24t \lambda \big( 17 993 981 + 4 203 006 t\lambda \big) \Big) \notag\\ 
&\quad+ 36e^{(9D_{\mathrm{rot}}+\lambda)t} \Big( -34 596 408 + t\lambda \big( 25 824 160 + t\lambda \big[-10 113 804 + t\lambda (3 160 364 + 2 259 805 t\lambda) \big] \big) \Big) \bigg) \notag\\
&\quad+ 2 D_{\mathrm{rot}}^{8} e^{5D_{\mathrm{rot}}t + 2\lambda t} \lambda^{4} \bigg(338 499 +28 503 520e^{5D_{\mathrm{rot}}t} + e^{8D_{\mathrm{rot}}t} \Big( 7 343 987 069 + 12t\lambda \big( 303 766 901 + 60 503 268 t\lambda \big) \Big) \notag\\
&\quad+ 72e^{(9D_{\mathrm{rot}}+\lambda)t} \Big( -102 400 404 + t\lambda \big( 95 240 100 + t\lambda \big[ -41 759 253 + t\lambda (13 470 521 + 3 160 080t\lambda) \big] \big) \Big) \bigg) \notag\\ 
&\quad+8 D_{\mathrm{rot}}^{9}e^{5D_{\mathrm{rot}}t + 2\lambda t} \lambda^{3} \bigg( 39 285 + 5 848 040e^{5D_{\mathrm{rot}}t} + e^{8D_{\mathrm{rot}}t} \Big( 1 723 053 355 + 6t\lambda \big( 210 883 577 + 53 532 888 t\lambda \big) \Big) \notag\\ 
&\quad+ 6 e^{(9D_{\mathrm{rot}}+\lambda)t} \Big( -288 156 780 + t\lambda \big( 388 150 188 + t\lambda \big[ -216 666 333 + 2t \lambda ( 41 892 261 + 5 957 900 t\lambda) \big] \big) \Big) \bigg)  \notag\\
&\quad+ D_{\mathrm{rot}}^{7}e^{5D_{\mathrm{rot}}t + 2\lambda t} \lambda^{5} \bigg( 946 971 +40 268 800e^{5D_{\mathrm{rot}}t} + e^{8D_{\mathrm{rot}}t} \Big( 6 089 723 909 + 24t \lambda \big( 105 467 359 +20 319 858 t\lambda \big) \Big) \notag\\ 
&\quad+ 36e^{(9D_{\mathrm{rot}}+\lambda)t} \Big( -170 303 880 t\lambda \big( 135 810 864 + t\lambda \big[-53 814 612 + t\lambda (16 348 576 + 6 538 035 t\lambda) \big] \big) \Big) \bigg) \Bigg),\notag\\
\end{align}
\begin{align}
\langle \Delta y(t)^4 \rangle &= 12D^{2}t^{2} + \frac{48 D D_{\mathrm{rot}} v^2 t (e^{-(D_{\mathrm{rot}}+\lambda)t} -1 + (D_{\mathrm{rot}}+\lambda)t )}{(D_{\mathrm{rot}} + \lambda)^{2} (4D_{\mathrm{rot}} + \lambda)} \notag\\ 
&\quad+ \frac{1}{30D_{\mathrm{rot}}(D_{\mathrm{rot}}+\lambda)^4 (4D_{\mathrm{rot}}+\lambda)^4 (9 D_{\mathrm{rot}}+\lambda)^2 (16D_{\mathrm{rot}}+\lambda)} \Bigg( \notag\\
&\quad e^{-14D_{\mathrm{rot}}t -3\lambda t} \Bigg[ -3 e^{5D_{\mathrm{rot}}t + 2\lambda t}(9-64e^{5D_{\mathrm{rot}}t} +55 e^{8D_{\mathrm{rot}}t})\lambda^8 + 29 859840D_{\mathrm{rot}}^{10} e^{14D_{\mathrm{rot}}t + 3\lambda t}t^2 \notag\\ 
&\quad+ 8D_{\mathrm{rot}}e^{5D_{\mathrm{rot}}t + 2\lambda t}\lambda^{7} \Big( -54 + 824e^{5D_{\mathrm{rot}}t} + 5e^{8D_{\mathrm{rot}}t}(-154+9\lambda t) \Big)  \notag\\ 
&\quad+ 207360 D_{\mathrm{rot}}^{9} e^{13D_{\mathrm{rot}}t + 2\lambda t}t \left(-240 + e^{(D_{\mathrm{rot}}+\lambda)t}(-540+401 \lambda t) \right) \notag\\
&\quad+ 2 D_{\mathrm{rot}}^{2} e^{5D_{\mathrm{rot}} +2\lambda t} \lambda^{6}\Big(-1377+42592e^{5D_{\mathrm{rot}}t} +5e^{8D_{\mathrm{rot}}t}(-8243 +1632 \lambda t) \Big) \notag\\ 
&\quad+ 4D_{\mathrm{rot}}^{3}e^{5D_{\mathrm{rot}}t + 2\lambda t}\lambda^{5} \Big( -2241 +129136e^{5D_{\mathrm{rot}}t} +5e^{8D_{\mathrm{rot}}t}(-21923 +14 268\lambda t) \notag\\
&\quad+ 360e^{(9D_{\mathrm{rot}} + \lambda)t} \big(-48+ \lambda t ( 12+ \lambda t ) \big) \Big)  \notag\\
&\quad+ D_{\mathrm{rot}}^{4} e^{5D_{\mathrm{rot}}t + 2\lambda t} \lambda^{4} \Big( -16011 + 1546816e^{5D_{\mathrm{rot}}t} +25e^{8D_{\mathrm{rot}}t} (-13885 + 97824 \lambda t) \notag\\ 
&\quad+ 2880e^{(D_{\mathrm{rot}}+r)t} \big( -411 +11 \lambda t (9 +2 \lambda t) \big) \Big)  \notag\\
&\quad+4 D_{\mathrm{rot}}^{5}e^{5D_{\mathrm{rot}}t + 2\lambda t} \lambda^{3} \Big( -3969 + 626384e^{5D_{\mathrm{rot}}t} + 5e^{8D_{\mathrm{rot}}t}(136877 +524130 \lambda t) \notag\\ 
&\quad+ 720e^{(9D_{\mathrm{rot}} + \lambda)t} \big( -1815 + \lambda t (153 +371 \lambda t) \big) \Big) \notag\\ 
&\quad+ 16D_{\mathrm{rot}}^{6} e^{5 D_{\mathrm{rot}}t + 2\lambda t} \lambda^{2} \Big( -513 +141028e^{5 D_{\mathrm{rot}}t} +5e^{8D_{\mathrm{rot}}t}(-143123 + 224 238 \lambda t) \notag\\ 
&\quad+ 180e^{9(D_{\mathrm{rot}}+\lambda)t} \big( 3195 + \lambda t (-5043 +3074 \lambda t) \big)  \Big)  \notag\\
&\quad+11520 D_{\mathrm{rot}}^{8}e^{10D_{\mathrm{rot}}t + 2\lambda t} \Big(18 -6e^{3D_{\mathrm{rot}}t}(2352 +1033 \lambda t) + e^{(4D_{\mathrm{rot}} + \lambda)t} \big( 14094 + \lambda t (-16389 + 7259 \lambda t) \big)  \Big)    \notag\\
&\quad+ 96 D_{\mathrm{rot}}^{7}e^{5 D_{\mathrm{rot}}t + 2\lambda t} \lambda  \Big( -18 + 11118e^{5D_{\mathrm{rot}}t} -10e^{8D_{\mathrm{rot}}t}(106347 + 12502 \lambda t) \notag\\
&\quad +15e^{(9D_{\mathrm{rot}} +\lambda)t } \big( 70158 + \lambda t (-63558 +26513 \lambda t) \big) \Big) \Bigg]v^{4} \Bigg).\notag\\
\end{align}
\end{subequations}

\section{Numerical methods \label{ap:numerics}}
\subsection{Eigenvalue problem}

To evaluate Eq.~\eqref{ISF_free}, we need the eigenvalues $a_{2m}(q)$ and eigenvectors $A_0^{2m}(q)$. Equation~\eqref{eigenval_rec} yields the matrix eigenvalue problem:
\begin{equation}
	 4p^2 A_{2p}^{2n} + q (A_{2p-2}^{2n}+A_{2p+2}^{2n}) = a_{2n}A_{2p}^{2n},
\end{equation} 

\noindent where the matrix is infinite, symmetric (but not Hermitian since $q$ is imaginary) and tridiagonal. This allows for an efficient numerical computation and, in practice, we truncate the order of the matrix to sufficiently high order such that the normalization for $t=0$ in Eq.~\eqref{ISF_free} is achieved. 

The numerical computations were made using the Julia programming language~\cite{Bezanson2017} and figures were produced with the package Makie.jl~\cite{Danisch2021}. 

\subsection{Stochastic simulations}
\label{stochastic_simulations}

To perform stochastic simulations, we discretize Eqs.~\eqref{stoch_r} and \eqref{stoch_theta} according to the Euler-Maruyama scheme:
\begin{subequations}
\begin{align}
\bm{r}(t+ \Delta t) &= v \bm{u}(\vartheta(t)) \Delta t + \sqrt{2D \Delta t} \bm{N}_\eta(0, 1), \\ 
\vartheta(t+ \Delta t) &= \sqrt{2D_{\mathrm{rot}} \Delta t} N_{\xi}(0, 1),
\end{align}
\end{subequations}
\noindent where $\Delta t = 10^{-3} \tau_{\mathrm{rot}}$ is the timestep, $\bm{N}_\eta(0, 1)$ and $N_{\xi}(0, 1)$ are independent, normally distributed random variable with zero mean and unit variance. Furthermore, the statistics are obtained by simulating trajectories for $10^5$ particles.
\twocolumngrid

\bibliographystyle{aipnum4-2}
\bibliography{bibliography3}

\end{document}